\documentstyle[prd,aps]{revtex}
\begin{document}

\title{
Head-on collisions of unequal mass black holes: close-limit
predictions}

\author{Zeferino Andrade and Richard H. Price}
\address{
Department of Physics, University of Utah, Salt Lake City, UT 84112}
\maketitle
\begin{abstract}
  The close-limit method has given approximations in excellent
  agreement with those of numerical relativity for collisions of equal
  mass black holes. We consider here colliding holes with unequal
  mass, for which numerical relativity results are not available.  We
  try to ask two questions: (i) Can we get approximate answers to
  astrophysical questions (ideal mass ratio for energy production,
  maximum recoil velocity, etc.), and (ii) can we better understand
  the limitations of approximation methods. There is some success in
  answering the first type of question, but more with the second,
  especially in connection with the issue of measures of the intrinsic
  mass of the colliding holes, and of the range of validity of the
  method.
\end{abstract}


\section*{I. Introduction and overview}

Recently much attention has been given to the problem of the
collision of two black holes. There are two main reasons for such
interest.  First, black hole collisions provide one of the most
powerful and interesting sources for possible detection by
gravitational wave observatories\cite{ligo}. Second, black hole
collisions are being intensively studied numerically by using
supercomputers to evolve Einstein's equations from initial data
representing two holes\cite{grandchallenge}.

The enormous difficulty of solving the equations numerically provides
motivation for approximation methods that, with little effort, can give
guidance to what results can be expected, what cases are interesting
for full numerical study, etc.
One method that has proved surprisingly successful in a range of tests
is the close-limit approximation (CLAP). This method applies to
initial value data representing holes which are initially close to
each other. If the holes are close enough then the horizon will
initially surround both holes, and the spacetime outside the horizon can
be considered to be a single perturbed hole. Strong nonsphericity
inside the initial horizon cannot affect the evolution of the fields
outside, and hence does not influence the generation of outgoing
radiation. The nonspherical perturbations outside the horizon can be
analyzed with the well developed techniques of linear perturbation
theory.

The method, applied to numerically or analytically generated initial
data was discussed by Abrahams and Price\cite{AP1}, and has been
applied to: (i) simple analytic initial data for the head-on collision
of momentarily stationary black holes\cite{price_pullin94,all,AP2},
(ii) numerically generated initial value data for holes which are
initially moving towards each other\cite{abrahams_cook94}, (iii)
analytic initial value solutions for holes which are initially moving
slowly towards each other\cite{baker_boost}, (iv) analytic initial
data for holes which have opposite inital angular momentum and are
initially momentarily stationary\cite{screw}.  Where comparisons with
full numerical results are available (all of the applications above,
except the last) the results of the CLAP method are found to be
remarkably successful, even when initial conditions would seem to
violate the assumptions underlying the approximations. This success
holds the promise of giving easy approximate answers about black hole
collisions.

All applications listed above have been to collisions of holes with
equal masses. (Those are essentially the only cases for which
comparisons have been available with fully numerical computations.)
Here we consider collisions of unequal mass holes, though comparisons
with numerical results cannot yet be made. There are two reasons we do
this. First, there are some interesting questions to which it is
better to have a very rough answer, even an uncertain answer (if the
nature of the uncertainty is kept in mind), than no answer at all. It
is interesting to see whether there are any surprises in predicted
radiation efficiency, in dependence on details of initial data, etc.
One interesting question applies {\em only} when the masses are
unequal: can gravitational radiation from the collision contain a
significant amount of linear momentum, so that the hole that forms
will recoil with an astrophysically significant proper velocity?

The second motivation is to look at the range of validity of CLAP
calculations. The ultimate index of validity (short of full numerical
relativity comparisons) is second-order perturbation calculations, and
work on this technique is well underway\cite{2ndorder,prep}. These
necessary calculations, however, are quite difficult (though far
easier than full numerical computations) and it is useful to look at
whether simple guidelines exist.  Our approach here is simultaneously
to use the CLAP method to look for astrophysical answers, and to use
the examples to gain deeper understanding of the method.

The nature of the questions being asked justifies avoiding unnecessary
complications and using the simplest initial data sets applicable. We
therefore limit our attention here to nonspinning holes which start
from rest and undergo a head-on collision. More specifically, we limit
our investigation to two momentarily stationary initial solutions.
Both initial geometries are conformally flat, and therefore are
completely determined by a conformal factor $\Phi^4$, where $\Phi$
solves the flat spacetime Laplace equation\cite{BowenYork}. The
simplest solution is that of Brill and Lindquist\cite{bl} (hereafter
``BL'') in which $\Phi$ has the form of the gravitational potential
for two Newtonian point masses.  Another solution, that of Misner and
Lindquist\cite{images,misner,lind63} (hereafter ``ML'') is more
complicated but has the very useful feature of an easily located
minimal area of the Einstein-Rosen bridges in the initial geometry.
It is also the initial geometry that has been used in almost all
numerical relativity studies of collisions of black holes.  For these
two sets of initial data we look at radiated energy as a function of
how close the initial holes are.  To characterize the separation of
the holes, we use in all cases the proper distance $L$, along the
symmetry axis, between the apparent horizons of each of the holes.

An important issue that arises is how to characterize the ``bare mass''
of each of the throats, i.e., how to assign an intrinsic mass to each
of the holes participating in the collision, a mass unaffected (in
some sense) by the presence of other nearby sources of gravitation.
There is a fairly natural choice of bare mass for the BL initial data,
but not for the ML initial data. We consider three candidates as bare
mass measures in the ML case. One of the more interesting conclusions
of this work is the importance of the choice of bare mass, and the
unphysical consequences of the wrong choice.

In Sec.~II we start by describing the BL and ML solutions and the
choices that can be made for bare mass. We then describe the CLAP
method. CLAP estimates of radiated energy are presented in Sec.~III
where it is seen that the choice of bare mass of one ML hole governs
even the qualitative nature of some answers.  In Sec.~IV results are
given for the recoil velocity of the final hole formed, due to the
emission of radiation.  In the results of Secs.~III and IV, the
usefulness of the results depend heavily on the range of validity of
the CLAP. In Sec.~V we look at a simple criterion for when linearized
theory should be applicable.  We find that the validity of this
criterion, which seems useful in the case of equal mass holes, does
not seem to extend to collisions in which the ratio of hole masses is
very small. Conclusions are presented in Sec.~VI. Details of several
calculations are presented in three appendices.

\section*{II. Approximation method and initial geometries}
\subsection*{Initial value solutions}
For simplicity we limit attention to time symmetric initial data.
There is then an initial hypersurface on which the extrinsic curvature
is zero so that initial value data for Einstein's equations consists
only of the three geometry of this hypersurface.  Also for simplicity,
we limit our consideration to three geometries that are conformally
flat.  This simplification is neither selected for or against by any
strong physical argument, but it leads to a very convenient
mathematical description, and is therefore used, e.g., in most
numerical relativity work.

We write  the 
conformally flat axisymmetric three metric  in the form
\begin{equation}\label{conflat}
ds^{2}=\Phi^{4}(R,\theta)(dR^{2}+R^{2}d\Omega^{2})\ ,
\end{equation}
where $d\Omega^2=d\theta^2+d\varphi^2\sin^{2}\theta$, and 
$R, \theta, \varphi$ are
spherical coordinates in the flat conformal space.  The Einstein
initial value equations then turn out to require only that 
$\Phi$ obey Laplace's equation
\begin{equation}\label{laplace}
\nabla^{2}\Phi=0\ ,
\end{equation}
where $\nabla^{2}$ is the Laplacian with respect to the flat  metric
$dR^{2}+R^{2}d\Omega^{2}$. 

It is necessary to find solutions corresponding to two initially
static black holes. To investigate the possible sensitivity of
radiation to the details of the initial value solution we consider two
different solutions of (\ref{laplace}).  Brill and Lindquist
investigated the solution to (\ref{laplace}) that is most immediately
apparent\cite{bl}, the solution with the form of the Newtonian
gravitational potential of two mass points:
\begin{equation}\label{blform}
\Phi=1+\frac{\alpha_{1}}{|\vec{R}-\vec{R}_{1}|}+
\frac{\alpha_{2}}{|\vec{R}-\vec{R}_{2}|}\ .
\end{equation}
Here all the vectors and their norms are defined in the flat three
dimensional space, with $\vec{R}$ being the position of an arbitrary
point in such space and $\vec{R}_{i}$ is the position of a point in
the flat space representing hole $i$.  Though (\ref{blform}) suggests
two pointlike solutions, there are in fact no physical singularities
corresponding to these points. Rather the three geometry near each
``point'' can be extended, through a throat, out to an asymptotically
flat space (an Einstein-Rosen bridge).  The complete BL three
geometry, then, includes three asymptotically flat regions. One is a
region with two throats connected, the region of ``our universe.'' The
other two regions contain one throat each corresponding to mass 1, and
to mass 2. (These distinctions are meaningful only when the two
throats are well separated in ``our'' universe; see \cite{bl}{\bf}).

The BL topology has a useful practical feature. Since hole 1 (for
example) has its own asymptotically flat region we can infer a ``bare
mass'' $m_1$ for hole 1 from the metric at large distances from the
throat. In this way we can compute bare masses $m_i$ for each of the
individual holes.  (In the limit that the holes are very far apart,
the bare masses are related to the parameters $\alpha_i$ in
(\ref{blform}) by $\alpha_i\approx m_i/2$, where we use $c=G=1$
units.)  
We can also compute the total mass $M$ in ``our'' universe,
the ADM mass of the two-hole spacetime.

Another solution to (\ref{laplace}), corresponding to two throats, of
the form
\begin{equation}\label{mlform}
\Phi=1+\sum_{n=1}^{\infty}(\frac{a_{n}}{|\vec{R}-\vec{d}_{n}|}+
\frac{b_{n}}{|\vec{R}-\vec{e}_{n}|})\ ,
\end{equation}
can be constructed by placing the ``masses'' $a_{n}$ and $b_{n}$ at
the locations $\vec{d}_{n}$ and $\vec{e}_{n}$ in the flat conformal
three geometry in a manner similar to the placement of electrical
charges in the problem of finding the electric potential distribution
in the region outside two charged conducting spheres. With an infinite
set of image ``masses'' it is possible to construct a three geometry
with two throats which open into two identical asymptotically flat
regions.  The isometry between the two regions takes the form of a
reflection through spheres at the minimal neck of each of the throats.
The details of the symmetrization procedure are given in Appendix B
(see also Lindquist\cite{lind63}).

For our purposes, the disadvantage of ML solutions is that there is no
immediate meaning that can be given to ``bare mass.''  We consider
three candidates for bare mass in ML solutions.
One candidate is the bare mass suggested by Lindquist in his study of
symmetrized initial value solutions\cite{lind63}. The ``Lindquist
mass'' sums the bare masses of all the images associated with one of
the throats (a divergent sum) then subtracts a Newtonian expression
for the binding energy due to the interaction of those images with
each other (another divergent sum). This definition of bare mass seems
not to have been used in recent work on the problem. Much more
commonly cited is a rather straightforward measure of the bare mass
that we shall call the ``area-mass.'' One takes the area $A_i$ of the
minimal throat on the initial hypersurface and computes a bare mass
from it as if it were an isolated Schwarzschild horizon:
$m_i=\sqrt{A_i/16\pi}$.
A third candidate for the bare mass is the ``Penrose
mass''\cite{penrose,tod,global}. This is a quasi-local definition of mass
interior to a 2-sphere that starts with a method, valid in linearized
theory, of extracting mass information from the Weyl tensor. This
method is then formally converted to curved spacetime. The method,
cannot be applied to all spacetimes, but it is always applicable to
axisymmetric cases. Some details of the computation of the Penrose 
mass are given in Appendix C.

\begin{figure}
\caption{For two holes with fixed Penrose mass $m_{iP}$, in the ratio
  $m_{2P}/m_{1P}=3/7$, the dependence of the Lindquist mass $m_{iL}$
  on separation is shown.}
\label{masscomp}
\end{figure}
Figure~\ref{masscomp} illustrates the issue of choice of bare mass in
the ML case with an example in which the ratio of Penrose bare masses
of the two holes is 7:3.  For each hole, the ratio of Lindquist mass
to Penrose mass is shown as a function of separation. For both the
heavy and the light holes, the Lindquist mass decreases relative to
the Penrose mass, but the decrease is more dramatic for the heavy
hole.  If the statement ``put the same two holes at different
separation'' means putting holes of fixed Penrose mass at different
separations, then these ``same'' two holes at small separation have
very different Lindquist masses, and a very different ratio of
Lindquist masses.  A similar comparison between Lindquist mass and
area-mass tells a very different story. If the $m_1/m_2$ ratio of
area-mass is 7:3 then, to considerable accuracy, the ratio of
Lindquist masses is also 7:3.  There is not a strict equivalence of
the two ratios. At very small separation there is a small deviation in
the ratios. More important, the value of the Lindquist mass and that
of the area-mass for a given hole are not the same when separations
are small, but even here the effects are small. This is shown in
Fig.~\ref{lindvarea} for two equal mass hole. At small distances
($L/M$ less than around 0.5) the Lindquist mass is discernibly larger
than the area-mass. By comparison, the difference in Penrose mass and
area-mass is 20 times larger.
In our energy and recoil estimates in the next two sections, no
difference can be seen in the results depending on which mass ratio is
held constant, the Lindquist mass or the area-mass. We shall,
therefore, present only the latter.
\begin{figure}
\caption{Difference between Lindquist and area mass as a function of separation
  when the holes are equal. Here $M$ is the total ADM mass of the spacetime.}
\label{lindvarea}
\end{figure}
\begin{figure}
\caption{For the bare mass $m_2$ of a hole, the ratio of the
Penrose measure of bare mass ($m_{2P}$) to the Lindquist measure
($m_{2L}$) is shown as a function of the ratio of (Lindquist) bare 
masses for two different separations $L/M$ (where $M$ is the
total ADM mass of the spacetime). The result shows that in the extreme
ratio limit different measures of bare mass agree.}
\label{fig:m2ratio}
\end{figure}

In the limit of extreme mass ratio (as well as in the limit of large
separation) the ambiguities of bare mass disappear.  When the limit
$m_2/m_1$ of the bare masses becomes very small, the ``particle''
limit for $m_2$, physical intuition suggests that any reasonable
definition of bare mass must agree with the proper mass of a point
particle perturbing the spacetime, and therefore all reasonable
definitions of bare mass will yield the same result. This is
illustrated in Fig.~\ref{fig:m2ratio}, which shows the ratio of
the Penrose and Lindquist measures of $m_2$, the bare mass of the less
massive hole, as the ratio of bare masses $m_2/m_1$ goes to zero. It
is clear that in this limit also all reasonable measures of $m_1$, the
bare mass of the more massive hole, will agree with each other and
with the ADM mass of the spacetime. As a check we have computed the
ratio of Penrose to Lindquist values of $m_1$ as $m_2/m_1$ decreases
and have verified that the ratio goes to unity (although not as
quickly as in Fig.~\ref{fig:m2ratio}).

\subsection*{The close approximation}
To apply the CLAP to the momentarily stationary axisymmetric space we
must rewrite (\ref{conflat}) in the form of a $t\approx$\,constant
slice of a Schwarzschild spacetime. This requires mapping the geometry
of (\ref{conflat}) onto a set of Schwarzschild-like coordinates
$r,\theta,\varphi$. We do this by transforming the radial coordinate
as if $R$ were the isotropic radial coordinate for the Schwarzschild
geometry:
\begin{equation}\label{rtrans}
R=(\sqrt{r}+\sqrt{r-2M})^2/4\ ,
\end{equation}
where $M$ is the total mass of the spacetime.
This transformation puts the spatial metric in the form
\begin{equation}\label{nearschw}
ds^{2}={\cal F}^4
\left(\frac{dr^{2}}{1-2M/r}+r^{2}d\Omega^{2}\right)\ ,
\end{equation}
where
\begin{equation}
{\cal F}\equiv
\frac{\Phi(R,\theta)}{1+M/2R}\ .
\end{equation}
The function $\Phi$ must satisfy (\ref{laplace}) and will have
singularities at the coordinate locations $R_i$ of ``mass points''(BL)
or ``images''(ML). Since $\Phi$ must approach the Schwarzschild
spatial metric $\Phi\rightarrow 1+2M/R$ as $R\rightarrow\infty$, we
can expand $\Phi$, for $R>\max{(R_i)}$, in Legendre polynomials $P_\ell$
as
\begin{equation}\label{expand}
\Phi=1+\frac{2M}{R}+\sum_{\ell=1}^\infty\gamma_{\ell}\left(\frac{M}{R}\right)^{\ell+1}P_\ell(\cos\theta)\ .
\end{equation}
We can eliminate the $\ell=1$
term from (\ref{expand}) by appropriate choice of coordinate origin. The
expression for ${\cal F}$, therefore, can be put into the form
\begin{equation}\label{asused}
{\cal F}=
1+\frac{1}{1+M/2R}\sum_{\ell=2}^\infty\gamma_{\ell}\left(\frac{M}{R}\right)^{\ell+1}P_\ell(\cos\theta)\ .
\end{equation}

For an initial value solution representing two black holes, the
coefficients $\gamma_\ell$ will contain a parameter describing the
separation of the holes. As the separation goes to zero, the geometry
approaches that of a $t=$\,constant slice of the Schwarzschild
geometry, and the $\gamma_\ell$ must therefore all approach zero.  The
close approximation consists of treating the separation of the holes
as a perturbation parameter and in (\ref{asused}) keeping only the
terms lowest order in the separation.  When we take the fourth power
of ${\cal F}$ the mixing of the $P_\ell$ gives us, in principle, very
complicated mixtures of contributions of different $\ell$ for each
final multipole of ${\cal F}^4$. The result in practice is much
simpler.  For both the BL and ML solutions (in fact for any
conformally flat metric (\ref{conflat}) for which (\ref{rtrans}) is
used) the $\gamma_\ell$ coefficients increase in the order of the
perturbation as $\ell$ increases and, as a result, for each
$\ell$-pole of ${\cal F}^4$ the contribution of lowest perturbative
order is due only to the $\gamma_\ell$ term, and the conformal factor
can be written
\begin{equation}\label{F4}
{\cal F}^4=
1+\frac{4}{1+M/2R}\sum_{\ell=2}^\infty\gamma_{\ell}\left(\frac{M}{R}\right)^{\ell+1}P_\ell(\cos\theta)\ .
\end{equation}
Equations (\ref{rtrans}), (\ref{nearschw}), and  (\ref{F4})
describe an initial perturbation of the Schwarzschild geometry.
Following the prescription of Moncrief\cite{moncrief74}, we can use this
perturbed initial solution to give us, for each $\ell$, the initial
value of a ``Zerilli function''\cite{zerilli} $\psi_\ell$. (We use the
normalization of Abrahams and Price\cite{AP1}, our $\psi_\ell$ being their
$Q^{+}_\ell$.)

This Zerilli function satisfies a simple wave equation and from  
its initial value it is simple to find the time evolved function
$\psi(t,r)$. Once this is known we can compute the gravitational energy 
radiated by the system of two black holes from
\begin{equation}\label{energy}
  \frac{E}{2M}=
  \frac{1}{32\pi}\sum_{\ell=2}^\infty \int dt|\frac{d\psi_\ell}{dt}|^{2}
  \frac{1}{2M}
\end{equation}
It can easily be shown  that the Zerilli functions corresponding to two
different solutions $A$ and $B$ of (\ref{laplace}) differ only in
amplitude, i.e. they are related through
\begin{equation}\label{twometric}
\psi_\ell^{B}(r,t)=\frac{\gamma_\ell^{B}}{\gamma_\ell^{A}}\psi_\ell^{A}(r,t)
\end{equation}
This means that if we know $\psi_\ell(r,t)$ for a given solution of
(\ref{laplace}), say the Misner equal mass solution, we can compute
the energy (Sec.~III) and the recoil velocity (Sec.~IV) for any other
conformally flat solution, directly.

\section*{III. Radiated energy}

We give here the radiated energy corresponding to a range of head on
collisions.  To choose a particular black hole collision, we must
first say whether the initial data is BL or ML. Second, we must
specify a ratio of hole bare masses, and third we must specify $L$ the
separation of the holes, i.e., the proper distance between the
apparent horizons.  Our results are plotted with the radiated energy
and the separation $L$ normalized by the sum of the bare masses, so that
the results can be interpreted as showing the change of radiation
with separation when the ``same'' two holes are moved closer together.

For ML holes we consider several different definitions of bare mass,
and in Fig.~\ref{fig:MLenergy} show the results for the Penrose
\begin{figure}
\caption{
For two ML holes, with bare mass ratio $F\equiv m_2/m_1$, the radiated
energy is shown as a function of the proper separation $L$ between the
apparent horizons. Results are shown for both the Penrose bare mass definition
and the Lindquist definition. For extreme mass ratios the choice of bare
mass is irrelevant, but for $F=0.75$ a qualitatively different dependence on
separation shows up.}
\label{fig:MLenergy}
\end{figure}
definition and the Lindquist (equivalently, the area) meaning of bare
mass.  To avoid cluttering the plot the results for $F=1$ have been
omitted. These $F=1$ results show little difference depending on the
choice of bare mass definition. The $F=1$ curves are only slightly
different for Penrose and Lindquist mass, and both are qualitatively
similar to the results for $F=0.75$, with Penrose mass.

The results in Fig.~\ref{fig:MLenergy} for $F=0.01$ show the
expected independence of bare mass definition. But they fail to show
another important feature that has appeared in a study of the particle
limit\cite{loustoprice}.  For particles that fall, starting from rest
at distance $L$ from a Schwarzschild hole, here is an anomalous
decrease of radiation with decreasing $L$. This is found only in the
range $L=5M$ to $7M$, where $M$ is the total ADM mass of the
spacetime. This unexpected reversal of the general trend is a
manifestation of the role played by the ``curvature potential'' (due
to strong field effects near the hole) which peaks at around $L=3M$.

For our $F=0.01$ results the ADM mass should be negligibly different
from $m_1+m_2$. So the plot of radiated energy should show this effect
in Fig.~\ref{fig:MLenergy}. The fact that it does not can be ascribed
to failure of the close limit.  We know in fact that this is
the case from the study of the particle limit\cite{loustoprice}. For
extreme mass ratios the close limit will work only for $L$ less than
about $2M$, and when it applies the radiation decreases with
decreasing $L$.

For equal mass holes there is no anomalous region. Radiated energy is
a monotonic function of initial separation. This can be interpreted to
mean that when the smaller hole is not at the particle limit it is
``too large'' to be situated in the narrow range of radii for which
the anomalous energy-separation behavior occurs. As we decrease the
bare mass ratio $F$ from unity, there must be some value at which the
anomalous region appears. The results in Fig.~\ref{fig:MLenergy} are
interesting in connection with this. For $F=1$ it is well established
that close limit results are reasonably accurate for values of $L/M$
around 4. It would suggest that the anomalous behavior
seen at rather small $L$ for the $F=0.75$ is not an artifact of the
close limit. The dramatic low-$L$ bump in that curve, of course, has
no counterpart in the corresponding Penrose curve, which suggests that
what we are seeing is an effect due to an anomalous choice of bare
mass. 
\begin{figure}
\caption{
For two BL holes, with bare mass ratio $F\equiv m_2/m_1$, the radiated
energy is shown as a function of the proper separation $L$ between the
apparent horizons.}
\label{fig:BLenergy}
\end{figure}
For BL collisions, as explained in the previous section, there is a
``favored'' definition of bare mass. It is straightforward in
principle and easy in practice to compute a reasonable mass of each
hole by going to large distances in the asymptotically flat region in
which that hole is the only throat.  For this definition of bare mass,
Fig.~\ref{fig:BLenergy} shows the radiated energy as a function of
$L$, with no anomalous behavior.  This supports the argument that the
low-$L$ bump, for the Lindquist $F=0.75$ curve in
Fig.~\ref{fig:MLenergy} is due to an unphysical choice of bare mass.

\section*{IV. Recoil velocity}
For unequal mass holes colliding along the $z$ axis there will be
momentum contained in the outgoing radiation due to the interaction of
multipoles of different $\ell$. Moncrief\cite{moncrief80} has
considered a similar effect in the radiation emitted by a collapsing
star.

The general expression for the rate at which  $z$ momentum, is radiated
is given by
\begin{equation}\label{genmom}
\frac{dP^{z}}{dt}=\frac{1}{16\pi}\sum_{l=2}^{\infty}\sqrt{\frac{(l-1)(l+3)}{(2l+1)(2l+3)}}\frac{d\psi_{l}}{dt}\frac{d\psi_{l+1}}{dt}\ .
\end{equation}
The mixture of $\ell=2$ and 3 is dominant in the sum (\ref{genmom}) when
the holes start close but the next mixture becomes more and more important
as they start farther and farther apart.
As a result of radiation emission, the final hole formed will
aquire a velocity (relative to the asymptotic frame in which 
the colliding holes were initially at rest). This recoil velocity
is
\begin{equation}
v^{z}=-\frac{1}{M}\int dt \frac{dP^{z}}{dt}.
\end{equation}
In terms of (\ref{genmom}) the recoil velocity is
\begin{eqnarray}\label{genv}
& &
v^{z}=
-\frac{1}{16\pi}\sum_{l=2}^\infty\sqrt{\frac{(l-1)(l+3)}{(2l+1)(2l+3)}}\times\nonumber\\
& &
\int\frac{d\psi_{l}}{dt}\frac{d\psi_{l+1}}{dt}\frac{1}{M}\,dt\ .
\end{eqnarray}
Now the Zerilli functions $\psi_\ell(t)$ and $\psi_{\ell+1}(t)$ must
be evolved numerically and the overlap integral in (\ref{genv}) must
be numerically computed.  The functions $\psi_\ell(t)$ and
$\psi_{\ell+1}(t)$ have a fixed form and the details of the collision
influence only the amplitude. The overlap integral in (\ref{genv})
therefore needs to be integrated only once. [See (\ref{twometric}).]

This recoil velocity is always negative. Since our convention is always 
to put the more massive of the colliding holes
on the positive z axis and the less massive on the negative side, this means
that the final hole moves in the direction from which the smaller mass 
approached.
Results are shown in Fig.~\ref{fig:BLrec} for the recoil velocity, in the 
case of BL initial data, as a function of separation.
\begin{figure}
\caption{Recoil velocity for BL holes. The recoil velocity of
  the final hole is shown as a function of initial separation for
  several  ratios of bare mass.}
\label{fig:BLrec}
\end{figure}
At the low values of $L$ at which the CLAP should be
reliable, the highest values of recoil velocity occur for $F\sim0.3$.
The recoil velocity results for ML initial conditions, shown in
Fig.~\ref{fig:MLpenrec} are roughly similar.  Mass ratios $F\sim0.1-0.3$
produce the highest recoil velocities.
\begin{figure}
\caption{For two ML 
  holes the recoil velocity of the final hole formed is shown as a
  function of initial separation, for several different values of $F$,
  the ratio of Penrose bare masses.}
\label{fig:MLpenrec}
\end{figure}

The figures both for ML and BL initial conditions include recoil
velocities of many hundreds of km/sec, proper velocities that would be
astrophysically interesting. The large values of $v^z$ correspond, of
course, to large initial separations. Whether Figs.~\ref{fig:BLrec}
and or \ref{fig:MLpenrec} actually contain results of astrophysical
interest depends on whether the CLAP fails at the separations which
predict large recoils. This question then, gives us a very specific
motivation for turning, in the next section to a consideration of the
range of applicability of  CLAP.

\section*{V. Range of validity of CLAP}

There is nothing inherent in the linear perturbation theory underlying
the CLAP to indicate how small the expansion parameter $\epsilon$ must
be to have answers accurate to, say, 10\%. In other words, we do not
immediately know ``how close is close enough.''
The only {\em a priori} justifiable way of finding the range of
validity (aside of course from full numerical solutions) is carry out
second-order perturbation theory, and to see at what $\epsilon$ the
first and second-order answers differ.  The formalism for higher order
computations has been developed\cite{2ndorder}, and successfully
applied to the equal mass case\cite{prep}. Higher order perturbation
theory, though much easier than full numerical solutions, is lengthy
and tedious. For this reason it is useful to look for easy rough
indices of validity of CLAP. 

To understand the importance of an index of validity one can look at
the application of CLAP to the collision of equal mass ML
holes\cite{price_pullin94,all}.  Here a comparison can be made with
the results of full nonlinear numerical analysis. Both methods predict
an increase of radiation with increasing initial separation. The
computed energies agree reasonably well (within a factor of $\sim2$)
for initial separations out to around $L/M\sim4$. The energy for infall
from extremely large distances is larger only by a factor $\sim2-3$.
It may be that these features are generally true: the magnitudes of
radiation quantities at the limit of CLAP validity may be within
better than an order of magnitude agreement with the large separation
limits.  In that case we can make good estimates for the large
separation case by taking the values at the CLAP limit.

An easy index for this limit has been suggested by Suen\cite{Suen}.
One can take the nonlinear initial data, extract perturbation
quantities and compute Moncrief's two even parity gauge-invariant
functions $q_1$ and $q_2$. For linearized data the linearized
Hamiltonian constraint gives $q_2=0$, so the magnitude of $q_2$ is an
indication of ``how nonlinear'' the initial data are, and therefore,
presumably, how accurate linearized computations are for the
evolution.  In Figs.~\ref{fig:misham1}, and \ref{fig:misham01}, we
plot the value of $q_2/q_1$ as a function of radius.  To make this
nonlinearity index more plausible we have modified it in two ways.
The factor of $(r/M-2)$ corrects for the divergence of $q_2$ at the
horizon.  If the mass ratio $m_2/m_2$ becomes very small, violations
of sphericity must scale as $m_2$. We would then find that violations
of nonlinearity were very small regardless how far apart the holes
start, and regardless how badly the CLAP fails. To take into account
the scaling of nonlinearities as $m_2$, we place a factor $F$ in the
denominator of the nonlinearity index. In the two figures we normalize
separation with the ADM mass $M$, rather than with bare masses. Since the
curves refer to a fixed set of masses, the ratio of $M$ to $m_1+m_2$ 
is fixed.

The first figure shows the result for equal masses
and ML initial geometry (see also\cite{AP1}). In this case we know,
from numerical relativity results, that the CLAP fails at around
$L/M\sim3-4$.  The results in Fig.~\ref{fig:misham1} show a dramatic
increase in the hamiltonian violation as $L/M$ increases beyond
$\sim2$ and this coincides, approximately, to the value of $L/M$ for
which CLAP starts to fail.

In Fig.~\ref{fig:misham01} we show the equivalent results for ML holes
with a ratio of Penrose mass $F=0.1$; in the previous section we saw
that ratios roughly around this value maximize the recoil velocity.
The results in this figure suggest that at $L/M=0.83$, CLAP must fail
badly. We know, from the particle limit study\cite{loustoprice} that
CLAP estimates are quite accurate (within 20\%) up to around
$L/M\sim1.5$ This implies that the CLAP method can work well even when
there is significant ``linearity violation'' in the initial data. It
also implies, unfortunately, that there may be no easy reliable way to
esitmate the validity of CLAP calculations.
 
Let us now tentatively accept, from the particle limit results that
for $F=0.1$ the CLAP method is valid out to around $L/M\sim1$, and
that the energy and linear momentum at this limit are, to order of
magnitude, as large as they would be for initially infinite separation
(as in the case of equal mass holes). The plots in
Figs.~\ref{fig:BLrec} and \ref{fig:MLpenrec} then tell us that recoil
velocities will always be well short of the several hundred km/sec
values that would be astrophysically interesting.

\begin{figure}
\caption{The linearized constraint violation for two equal mass ML
  holes, $F=1$ according with Penrose, is shown for several values of initial separation. Here $r^*$
is the coordinate $r+2M\ln{(r/2M-1)}$ for the Schwarzschild geometry.}
\label{fig:misham1}
\end{figure}

\begin{figure}
\caption{The linearized constraint violation for ML holes with
a ratio $F=0.1$ of the Penrose masses.}
\label{fig:misham01}
\end{figure}

\section*{VI. Conclusions}
The application of the CLAP method to collisions of unequal mass holes
has provided one very tentative conclusion: The linear momentum
generated in the collisions will be far too small to produce
astrophysically interesting recoil velocity of the final hole.

The study has been much more successful in raising questions and in
uncovering difficulties. The most important difficulty is the choice
of the bare mass of the holes participating in a collision.  It should
be understood that this is not an issue specific to CLAP estimates.
Rather, CLAP estimates were used to probe it. The general problem is:
How do we best characterize an individual hole when it is interacting
strongly with another gravitational source? And this issue is part of
the broader question:
How do we set up initial data to represent a black hole configuration?

Astrophysical models, or Newtonian physics applied to the early
dynamics, can give us an initial configuration in terms of simple
physical parameters (mass of the holes, separation,...). To turn this
into appropriate initial data for numerical relativity we must know
the correct general relativistic interpretation of the classical
picture. This will be of crucial importance to the use of numerical
relativity to study three dimensional black hole collisions. Codes to
evolve black holes tend to be unstable. It is therefore important
to apply the codes ``at the last possible moment,'' i.e., only to
follow the last orbit or last few orbits of black hole coalescence. It
will be necessary then for those codes to begin with initial data for
holes which are already interacting fairly strongly and it is in just
this case that there is the most ambiguity in translating the
Newtonian concept of the mass of a hole, into a parameter of the
relativistic initial data.

We have considered three measures of bare mass. The ``area-mass,'' and
the ``Lindquist mass'' are well defined only for ML data, for which
there are symmetric Einstein-Rosen bridges. The ``Penrose'' mass
measure (though not generally applicable) can be computed for either
ML or BL initial data (and in the latter case agrees with the obvious
choice of bare mass in the asymptotically flat region of a throat).
We have found very mixed messages about the physical meaningfulness of
the various mass measures when the holes are initially close. One
somewhat surprising message is that the area-mass and the Lindquist
mass of each throat are numerically quite close. They are even closer
if we make a more subtle comparison: the ratio of area/Lindquist
masses of two holes, and the ratio of Penrose masses. This agreement
is not a manifestation of some general necessity of all mass measures
to agree. The Penrose mass and the area/Lindquist mass are markedly
different for close separations.

The agreement of area-mass and Lindquist mass is unexpected, because
they are based on such different criteria.  The area-mass connects
mass to the area of the minimum section of the Einstein-Rosen bridge,
just as if it were an isolated Schwarzschild hole. It takes no direct
account of the presence of a second throat. The Lindquist mass, by
very sharp contrast, uses asymptotic masses of ``images'' used to form
a single hole and subtracts an expression for binding energy due to
gravitational interaction of the images. The binding energy is
computed by Newtonian physics applied in the conformally flat space
underlying the initial data. (Both the sum of asymptotic masses and
the binding energy are divergent, but the sum is not.)  It is often
the case that when two very different ways of measuring a physical
quantity agree it is taken as good evidence that the measurement is
valid. This would suggest that we take seriously the area/Lindquist
bare mass, perhaps for a wider class of problems.  But the strange
behavior seen in Fig.~\ref{fig:MLenergy} for mass ratio $F=0.75$
suggests even more strongly that this bare mass measure can be
misleading.

\section*{ Acknowledgments}

We would like to thank Carlos Lousto for many useful conversations,
and also Madhavan Varadarajan for discussions of Penrose mass.  We
thank Carleton Detar for help with computational aspects of the work.
ZA was supported by a PhD grant under PRAXIS XXI administered by JNICT
(Portugal). This work was partially supported by NSF grant
PHY-9507719.

\appendix
\section{The Brill-Lindquist solution for two black holes}
In this appendix we evaluate the coefficients $\gamma_\ell$ in the expansion 
(\ref{expand}) of the 
Brill-Lindquist solution (\ref{blform}).

When $|\vec{R}|>|\vec{R}_{1}|$ and $|\vec{R}|>|\vec{R}_{2}|$ the expansion
of (\ref{blform}) in Legendre polynomials gives
\begin{equation}
\gamma_\ell=[(\frac{R_{1}}{M})^{l}\frac{\alpha_{1}}{M}+
(\frac{R_{2}}{M})^{l}\frac{\alpha_{2}}{M}]\ .
\end{equation}
We can completely characterize the two holes by two parameters in the
flat space: the ``distance'' $z_{0}=|\vec{R}_{2}-\vec{R}_{1}|$ and the
``mass ratio'' $C=\alpha_{2}/\alpha_{1}$.  Choosing the origin of
coordinates in this space at the ficticious center of mass, we can
rewrite $\gamma_\ell$ as
\begin{equation}
\gamma_\ell=\frac{\alpha_{1}}{M}(\frac{z_{0}}{M})^{l}\frac{1}{(1+C)^{l}}[C^{l}+(
-1)^{l}C] \ .
\end{equation}
where we chose the convention that hole 1 is on the positive $z$ axis,
hole 2 on the negative side and $0<C\leq 1$.  Since the total mass of
the spacetime at the moment of time symmetry is
\begin{equation}
M=2(\alpha_{1}+\alpha_{2})\ ,
\end{equation}
i.e., 
\begin{equation}
\frac{\alpha_{1}}{M}=\frac{1}{2(1+C)}\ ,
\end{equation}
we obtain
\begin{equation}
\gamma_\ell=(\frac{z_{0}}{M})^{l}\frac{1}{2(1+C)^{l+1}}[C^{l}+(-1)^{l}C]\ .
\end{equation}
The two parameters $z_{0}$ and $C$ do not have a direct
physical meaning. We therefore introduce the ratio of the bare masses of 
each black hole, $F=m_{2}/m_{1}$, and the separation between the apparent 
horizons, L, 
measured along the axis of symmetry and we express $z_{0}$ and $C$ (and hence 
$\gamma_{l}$) in terms of them. 

If we choose, for simplicity, the origin of coordinates to be at singularity 1,then
$L$, the proper distance between apparent horizons is
\begin{equation}\label{tamanho}
L=\int_{z_{1}}^{z_{2}}[1+\frac{1}{2(1+C)}(\frac{1}{z}+\frac{C}{z_{0}-z})]^{2}dz\
 .
\end{equation}
Here $z_{1}$ and $z_{2}$ are the $z$-axis intersections of the
apparent horizons surrounding holes $1$ and $2$, respectively.  To
find $z_1,z_2$ we numerically integrated the system of ODEs that
determine all the extremal two dimensional surfaces of the BL
solution (see \cite{bishop} for details), and we searched
along the segment of the $z$ axis between the positions of the two
holes (i.e. between $z=0$ and $z=z_{0}$) for the critical values $z_1$ and $z_2$ at which  the extremal surfaces
are closed. 
The bare masses of hole 1 and hole 2 are\cite{bl}:
\begin{eqnarray}
& &
m_{1}=2\alpha_{1}(1+\frac{\alpha_{2}}{z_{0}})\\
& &
m_{2}=2\alpha_{2}(1+\frac{\alpha_{1}}{z_{0}})\ ,
\end{eqnarray}
and hence 
\begin{equation}\label{mratio}
F=\frac{m_{2}}{m_{1}}=\frac{1+\frac{1}{2(1+C)}\frac{M}{z_{0}}}
{1+\frac{C}{2(1+C)}\frac{M}{z_{0}}}C\ .
\end{equation}
Solving for $C$ in terms of $F$ and $z_{0}$ we obtain
\begin{equation}\label{ratfalse}
C=\frac{(F-1)(2+\frac{M}{z_{0}})+\sqrt{(1-F)^{2}(2+\frac{M}{z_{0}})^{2}+16F}}{4}
\ .
\end{equation}
Typically to characterize a BL solution, we choose an $F$ and a set of $z_{0}$
values. Then from (\ref{ratfalse}) we obtain the corresponding set of $C$
values and from (\ref{tamanho}) the $L$ value corresponding to each pair
$(C,z_{0})$.

\section{ The Misner-Lindquist solution for two black holes}

In this appendix we present the ML solution and some additional
details relevant to the CLAP of this solution. We will follow closely
Lindquist\cite{lind63} who derived such a solution in implicit
form. Our notation is also the same as his with minor changes.

\subsection{The solution}
The conformal factor can be written in the explicit form (\ref{mlform})
\begin{equation}\label{mlfactor}
\Phi=1+\sum_{n=1}^{\infty}(\frac{a_{n}}{|\vec{x}-\vec{d}_{n}|}+
\frac{b_{n}}{|\vec{x}-\vec{e}_{n}|})
\end{equation}
Where $\vec{d}_{1}$ is the position of sphere 1, of radius a and
$\vec{e}_{1}$ the position of sphere 2, of radius b, relative to a
given origin in the flat space (the spheres in the flat space
correspond to the throats in the three geometry) and $\vec{d}_{n}$
are the positions of the image poles of $\vec{d}_1$, $\vec{e}_{n}$ are
the positions of the image poles of $\vec{e}_{1}$ with $a_{n}$ and
$b_{n}$ being the corresponding weights. These coefficients obey
the recursion relations
\begin{itemize}
\item {\bf If n is even}
\begin{eqnarray*}
& &
d_{n}=e_{1}-\frac{b^{2}}{e_{1}+d_{n-1}}\\
& &
a_{n}=\frac{b}{e_{1}+d_{n-1}}a_{n-1}\\
& &
e_{n}=d_{1}-\frac{a^{2}}{d_{1}+e_{n-1}}\\
& &
b_{n}=\frac{a}{d_{1}+e_{n-1}}b_{n-1}
\end{eqnarray*}
\item {\bf If n is odd} ($n\geq3$)
\begin{eqnarray*}
& &
d_{n}=d_{1}-\frac{a^{2}}{d_{1}+d_{n-1}}\\
& &
a_{1}=a\\
& &
a_{n}=\frac{a}{d_{1}+d_{n-1}}a_{n-1}\\
& &
e_{n}=e_{1}-\frac{b^{2}}{e_{1}+e_{n-1}}\\
& &
b_{1}=b\\
& &
b_{n}=\frac{b}{e_{1}+e_{n-1}}b_{n-1}
\end{eqnarray*}
\end{itemize}
whose solution is (see for example Smythe\cite{smythe})
\begin{itemize}
\item {\bf If n is even}
\begin{mathletters}\label{generala}
\begin{eqnarray}
& &
d_{n}=e_{1}-c(1-\frac{ba\sinh(n+2)\mu_{0}+a^{2}\sinh n\mu_{0}}{c^{2}\sinh n\mu_{0}})\\ 
& &
a_{n}=\frac{ab\sinh2\mu_{0}}{c\sinh n\mu_{0}}\\
& &
e_{n}=d_{1}-c(1-\frac{ba\sinh(n+2)\mu_{0}+b^{2}\sinh n\mu_{0}}{c^{2}\sinh n\mu_{0}})\\ 
& &
b_{n}=\frac{ab\sinh 2\mu_{0}}{c\sinh n\mu_{0}}
\end{eqnarray}
\end{mathletters}
\item {\bf If n is odd}
\begin{mathletters}\label{generalb}
\begin{eqnarray}
& &
d_{n}=
d_{1}-c(1-\frac{\sinh(n+1)\mu_{0}}{\sinh(n+1)\mu_{0}+\frac{a}{b}\sinh(n-1)\mu_{0}})\\
& &
a_{n}=\frac{ab\sinh2\mu_{0}}{b\sinh (n+1)\mu_{0} +a\sinh(n-1)\mu_{0}}\\
& &
e_{n}=
e_{1}-c(1-\frac{\sinh(n+1)\mu_{0}}{\sinh(n+1)\mu_{0}+\frac{b}{a}\sinh(n-1)\mu_{0}})\\
& &
b_{n}=\frac{ab\sinh2\mu_{0}}{a\sinh(n+1)\mu_{0}+b\sinh(n-1)\mu_{0}}
\end{eqnarray}
\end{mathletters}
\end{itemize}
where $c$ is the distance (in the flat space) between the centers of the two
spheres
\begin{equation}
c=d_{1}+e_{1}
\end{equation}\label{mmm}
and $\mu_{0}$ is given by
\begin{equation}
\cosh2\mu_{0}=\frac{c^{2}-a^{2}-b^{2}}{2ab}\ .
\end{equation}

\subsection{The choice of parameters}
In the background three dimensional Euclidean space there are two
natural dimensionless parameters: the ratio $C=b/a$ of the radii $a$
and $b$ of the spheres 1 and 2, and the ratio of the distance between
the two throats (i.e. the distance between the centers of the spheres
$c=d_{1}+e_{1}$) to the radius of one of the throats say $a$:
$D=c/a$. We will always restrict attention to the cases $0<C\leq 1$
i.e. hole 1 on the positive $z$ axis is greater or equal to hole 2 on
the negative $z$ axis. We must also have $D>1+C$ and from (\ref{mmm})
\begin{equation}
\mu_{0}=\frac{1}{2}\cosh^{-1}[\frac{D^{2}-1-C^{2}}{2C}]
\end{equation}
These two parameters, like $C$ and $z_{0}/M$ in the BL solution,
completely characterize the Misner-Lindquist solution. As in the BL solution,
however, they do not have a direct physical meaning and we must look
to other way of parameterizing the ML solution. One parameter which
can be defined in a natural way in the ML solution, the distance
between the two holes defined by the length of the geodesic which
threads through their corresponding Einstein-Rosen bridges, $L$, is
discussed in the first part of this section. Another physical
parameter is the ratio between the ``bare'' masses of each ML hole,
$F=m_{2}/m_{1}$.  As discussed above, we lack a unique, well defined,
notion of bare mass of each hole. The Lindquist masses presented on the 
second part are just a possibility. Once we have a prescription that gives the masses we can parametrize the ML solution
either by F and $L/M$ or by F and $L/(m_{1}+m_{2})$.

Due to the difficulty of obtaining $C$ and $D$ as functions of these
physical parameters, we will instead derive all the quantities in
terms of $C$ and $D$. 

As we will see the restriction $D>1+C$, equivalent to $\mu_{0}>0$, is required
in order to obtain convergence of the different series which appear in the next sections. 

\subsubsection{Invariant distance of separation}
Instead of using the distance between the throats (i.e. the parameter
$D$ defined above) as measured in the Euclidean space, it makes more
sense physically to use the invariant separation distance between the
two holes defined as the length of the shortest closed path which
threads through their corresponding E-R bridges. Such a curve is a
geodesic of the initial slice. To evaluate its length it is convenient
to work in bispherical coordinates instead of cartesian ones.

In these coordinates the metric of the initial surface is
\begin{equation}
ds^{2}=\Phi^{4}(\frac{f}{\cosh\mu-\cos\eta})^{2}[d\mu^{2}+d\eta^{2}+
\sin^{2}\eta d\varphi^{2}]\ .
\end{equation}
We introduce
\begin{mathletters}\label{bisph}
\begin{eqnarray}
& &
a=f{\rm csch}\,\mu_{1}\\
& &
d_{1}=f\coth\mu_{1}\\
& &
b=f{\rm csch}\,\mu_{2}\\
& &
e_{1}=f\coth\mu_{2}
\end{eqnarray}
\end{mathletters}
where
\begin{eqnarray}
& &
\mu_{1}+\mu_{2}=2\mu_{0}\\
& &
\mu_{2}=\sinh^{-1}\frac{\sinh2\mu_{0}}{D}\\
& &
f=b\sinh\mu_{2}\ .
\end{eqnarray}
Substituting (\ref{bisph}) in (\ref{generala}) and (\ref{generalb}) we
obtain a convenient form for the coefficients of $\Phi$ in
(\ref{mlfactor}). (Compare with (4.7) in Lindquist\cite{lind63} when
the two black holes are equal.)
\begin{itemize}
\item {\bf If n is even}
\begin{mathletters}\label{speciala}
\begin{eqnarray}
& &
a_{n}=b_{n}=f{\rm csch}\, n\mu_{0}\\
& &
d_{n}=e_{n}=f\coth n\mu_{0}
\end{eqnarray}
\end{mathletters}
\item {\bf If n is odd}
\begin{mathletters}\label{specialb}
\begin{eqnarray}
& &
a_{n}=f{\rm csch}\, [(n+1)\mu_{0}-\mu_{2}]\\
& &
b_{n}=f{\rm csch}\, [(n-1)\mu_{0}+\mu_{2}]\\
& &
d_{n}=f\coth [(n+1)\mu_{0}-\mu_{2}]\\
& &
e_{n}=f\coth[(n-1)\mu_{0}+\mu_{2}]\ .
\end{eqnarray}
\end{mathletters}
\end{itemize}
and the conformal factor becomes
\begin{eqnarray}
& & \nonumber
\Phi=\sum_{n=-\infty}^{\infty}\frac{1}{\sqrt{\cosh(\mu+4n\mu_{0})-\cos\eta}}+\\
& &
\frac{1}{\sqrt{\cosh[\mu+4n\mu_{0}+2\mu_{2}]-\cos\eta}}(\cosh\mu-\cos\eta)^{\frac{1}{2}}
\end{eqnarray}
which converges if and only if $\mu_{0}>0$.

We note that this parametrization implies a specific choice of origin
in the Euclidean space, however the geodesic distance between the two
throats is the same independent of the choice of origin. It is thus
just a matter of convenience.  The equations of the throats
$x^{2}+y^{2}+(z-d_{1})^2=a^{2}$ and $x^{2}+y^{2}+(z+e_{1})^{2}=b^{2}$
become $\mu=\mu_{1}$ and $\mu=-\mu_{2}$ respectively.

The geodesic of interest is $\varphi=0$, $\eta=\pi$. Therefore the distance of separation between the holes is
\begin{eqnarray}
& &\nonumber
L=\int_{-\mu_{2}}^{\mu_{1}}\sqrt{g_{\mu\mu}(\mu,\pi)}d\mu=\\
& &
f\int_{-\mu_{2}}^{\mu_{1}}\Phi^{2}(\mu,\pi)
\frac{1}{\cosh\mu+1}d\mu\ .
\end{eqnarray}
or explicitly
\begin{eqnarray}
& &\nonumber
L=2f\left\{1+\frac{\mu_{2}}{\sinh(\mu_{2})}+\sum_{n=1}^{\infty}[\frac{4n\mu_{0}}{\sinh(2n\mu_{0})}\right .+\\
& &
\left.\frac{2n\mu_{0}-\mu_{2}}{\sinh(2n\mu_{0}-\mu_{2})}+\frac{2n\mu_{0}+\mu_{2}}
{\sinh(2n\mu_{0}+\mu_{2})}]\right\}
\end{eqnarray}
Again this series converges if and only if $\mu_{0}>0$.

\subsubsection{The Lindquist mass}
 The bare masses of the  holes, according to Lindquist\cite{lind63}, are
\begin{eqnarray}
& &
m_{1}=2\sum_{n=1}^{\infty}[a_{2n-1}+b_{2n}+\sum_{m=1}^{\infty}
\frac{a_{2n-1}a_{2m}}{|\vec{d}_{2n-1}-\vec{d}_{2m}|}+\nonumber\\
& &
\frac{a_{2n-1}b_{2m-1}}{|\vec{d}_{2n-1}-\vec{e}_{2m-1}|}
+\frac{b_{2n}a_{2m}}{|\vec{e}_{2n}-\vec{d}_{2m}|}+\frac{b_{2n}b_{2m-1}}{|\vec{e}_{2n}-\vec{e}_{2m-1}|}]
\end{eqnarray}
\begin{eqnarray}
& &
m_{2}=2\sum_{n=1}^{\infty}[a_{2n}+b_{2n-1}+\sum_{m=1}^{\infty}
\frac{a_{2n}a_{2m-1}}{|\vec{d}_{2n}-\vec{d}_{2m-1}|}+\nonumber\\
& &
\frac{a_{2n}b_{2m}}{|\vec{d}_{2n}-\vec{e}_{2m}|}
+\frac{b_{2n-1}a_{2m-1}}{|\vec{e}_{2n-1}-\vec{d}_{2m-1}|}+\frac{b_{2n-1}b_{2m}}{|\vec{e}_{2n-1}-\vec{e}_{2m}|}]\ .
\end{eqnarray}
which using the coefficients in (\ref{speciala}) and (\ref{specialb}) become
\begin{eqnarray}
& &
m_{1}=2f\sum_{n=1}^{\infty}n\{\frac{2}{\sinh 2n\mu_{0}}+\frac{1}{\sinh (2n\mu_{0}-\mu_{2})}+\nonumber\\
& &
\frac{1}{\sinh(2n\mu_{0}+\mu_{2})}\}
\end{eqnarray}
\begin{eqnarray}
& &
m_{2}=2f\sum_{n=1}^{\infty}n\{\frac{2}{\sinh2n\mu_{0}}+\frac{1}{\sinh[2(n+1)\mu_{0}-\mu_{2}]}+\nonumber\\
& &
\frac{1}{\sinh[2(n-1)\mu_{0}+\mu_{2}]}\}
\end{eqnarray}
Both series converge if and only if $\mu_{0}>0$.

\subsection{The choice of origin}
In order to determine the energy and the recoil velocity we need to
choose the origin in such a way that the dipole term in (\ref{expand})
vanishes. This requires that the origin  be located at the
``center of mass'' (CM) of the ``masses'' $a_{n}$, $b_{n}$,
\begin{equation}
\sum_{n=1}^\infty a_{n}\vec{d}_{nCM}+b_{n}\vec{e}_{nCM}=0\ .
\end{equation}
(The $a_{n}$ and $b_{n}$ do not depend on the choice of origin). One systematic way of determining the $\vec{d}_{nCM}$, $\vec{e}_{nCM}$ is the following

\begin{itemize}
\item We choose an arbitrary origin, for example the one for which bispherical
coordinates can be introduced and pick the corresponding $d_{n}$ and $e_{n}$
in (\ref{speciala}) and (\ref{specialb}).
\item We next find the center of mass position using
\begin{equation}
z_{CM}=\frac{\sum_{n=1}-a_{2n}d_{2n}+a_{2n-1}d_{2n-1}+
b_{2n}e_{2n}-b_{2n-1}e_{2n-1}}{\sum_{n=1}a_{n}+b_{n}}\ .
\end{equation}

\item Finally we determine the position of a given image relatively to the 
CM, using the formula
\begin{equation}
\vec{z}_{iCM}=\vec{z}_{i}-\vec{z}_{CM}
\end{equation}
or in scalar form
\begin{mathletters}\label{cmas}
\begin{eqnarray}
& &
d_{2nCM}=-d_{2n}-z_{CM}\\
& &
e_{2n+1CM}=-e_{2n+1}-z_{CM}\\
& &
e_{2nCM}=e_{2n}-z_{CM}\\
& &
d_{2n+1CM}=d_{2n+1}-z_{CM}
\end{eqnarray}
\end{mathletters}
\end{itemize}

\subsection{The $\gamma_\ell$ coefficients}

Comparing (\ref{mlform}) with (\ref{expand}), we obtain the coefficients $\gamma_\ell$
\begin{equation}
\gamma_\ell=\sum_{n=1}^\infty [\frac{a_{n}}{M}(\frac{d_{nCM}}{M})^\ell+\frac{b_{n}}{M}(\frac{e_{nCM}}{M})^\ell]\ .
\end{equation}

\subsection{The mass of the system}
To determine the $\gamma_\ell$ above we need to know a dimensionless
parameter, say the radius of sphere 2, b, per unit of mass. This can easily
be done by expressing the mass in terms of the coefficients in the conformal
factor. The ADM mass of the spacetime at the moment of time symmetry
is the coefficient of 
$1/r$ in an expansion of $ds^{2}$ in inverse powers of r. The result
is
\begin{equation}\label{totalmass}
M=2\sum_{n=1}a_{n}+b_{n}\ .
\end{equation}
If the explicit expressions in (\ref{generala}) and (\ref{generalb})
are used, this becomes
\begin{eqnarray}
& &
M=2f\sum_{n=1}^{\infty}\{\frac{2}{\sinh2n\mu_{0}}+\frac{1}{\sinh(2n\mu_{0}-\mu_{2})}+\nonumber\\
& &
\frac{1}{\sinh[2(n-1)\mu_{0}+\mu_{2}]}\}
\end{eqnarray}
from which we can express $b$ in terms of $C$ and $D$. We note that once more
the series converges if and only if $\mu_{0}>0$.

\section{The Penrose quasi-local mass of one ML black hole}
In Penrose's approach a complex quantity $A_{\alpha\beta}$, 
the momentum-angular 
momentum twistor of the source inside a 2-surface S ({\it with topology $S^
{2}$}) is defined and the
total mass threading through S is\cite{penrose}
\begin{equation}
m^{2}_{P}=-\textstyle{\frac{1}{2}}A_{\alpha\beta}\overline{A}^{\alpha\beta}\ .
\end{equation}

Certain problems with this definition remain. In order to evaluate this mass 
we
need a Hermitian ``norm'' defined for the surface S as a whole. This norm can
only be defined unambiguosly if S can be embedded in some conformally flat
space-time in such a way that both its intrinsic geometry (induced metric) and
the quantities characterizing its extrinsic curvature are unaltered from their
values when S is embedded in the true spacetime. 
However this is not a difficulty for the ML solution (and BL solution), since
 any time symmetric {\it conformally flat} hypersurface can be
embedded in a conformally flat 4-space as a surface of constant $t$ with the
same intrinsic and extrinsic geometries and hence also all 2-surfaces lying on this hypersurface. 

In fact Tod\cite{tod} has evaluated the Penrose quasi-local mass associated with 
a 2-surface S obeying the restriction above, including the time symmetric 
case.
He showed that in the BL case, the Penrose mass of each hole
coincides with the ADM mass and that for both BL and ML solutions 
the total Penrose mass of the spacetime at the instant of time symmetry is 
the same as the total ADM mass. He also derived the mass of each
ML hole, when the masses are equal.

 In this 
appendix we use Tod's results to derive the Penrose mass of each ML black hole 
when their masses are unequal. We will use a signature $(+---)$.

Let {\bf t} denote the unit timelike vector normal to the time
symmetric 3-surface $\Sigma$.  The Penrose mass ``enclosed'' by the
two surface S lying in $\Sigma$ with normal {\bf n} is
\begin{equation}
m^{2}_{P}=P_{a}\overline{P}^{a}-\textstyle{\frac{1}{2}}\overline{\lambda}^{AB}\mu_{AB}-
\textstyle{\frac{1}{2}}
\lambda^{A'B'}\overline{\mu}_{A'B'}
\end{equation}
which we can rewrite in terms of spinors as ( see Tod\cite{tod}))
\begin{eqnarray}
& &
m^{2}_{P}=-2t_{A'(B}P^{A'}_{A)}t^{C'(B}\overline{P}_{C'}^{A)}+
P^{b}t_{b}\overline{(P^{b}t_{b})}+\nonumber\\
& &
\overline{\lambda}^{AB}t^{A'}_{B}t^{C}_{A'}\mu_{AC}+
\lambda^{A'B'}t^{A}_{B'}t^{C'}_{A}\overline\mu_{A'B'}
\end{eqnarray}
where
\begin{equation}
\mu_{AB}t^{B}_{A'}=\frac{-i}{2\pi}\int\{2(\vec{n}.\nabla\Phi)\nabla\Phi
-(\nabla\Phi)^{2}\vec{n}\}dS
\end{equation}
\begin{eqnarray}
& &
t^{a}P_{a}=\frac{1}{2\pi}\int\{-\Phi\vec{n}.\nabla\Phi-2(\vec{n}.\nabla\Phi)
(\vec{R}.\nabla\Phi)+\nonumber\\
& &
(\nabla\Phi)^{2}(\vec{n}.\vec{R})\}dS
\end{eqnarray}
\begin{eqnarray}
& &
t_{A'(B}P^{A'}_{A)}=\frac{1}{2\pi}\int\{\Phi(\nabla\Phi\times\vec{n})+
(\nabla\Phi)^{2}\vec{n}\times\vec{R}-\nonumber\\
& &
2(\nabla\Phi.\vec{n})(\nabla\Phi\times
\vec{R})\}dS
\end{eqnarray}
\begin{eqnarray}
& &
t^{A}_{B'}\lambda^{A'B'}=\frac{i}{2\pi}\int\{\vec{n}(\frac{1}{2}\Phi^{2}
+\Phi(\vec{R}.\nabla\Phi)+\nonumber\\
& &
\frac{1}{2}R^{2}(\nabla\Phi)^{2})
+\vec{R}(2(\vec{n}.\nabla\Phi)(\vec{R}.\nabla\Phi)
-(\vec{n}.\vec{R})
(\nabla\Phi)^{2}+\nonumber\\
& &
\Phi\vec{n}.\nabla\Phi)
-\nabla\Phi(\Phi(\vec{n}.\vec{R})+R^{2}(\vec{n}.\nabla{\Phi})\}dS
\end{eqnarray}
with all expressions on the right hand side written in terms of a three vector notation on flat
three-space. To evaluate the integrals we transform them to  integrals over
a volume spanned by S
\begin{equation}
\mu_{AB}t^{B}_{A'}=-\frac{i}{\pi}\int \nabla\Phi\nabla^{2}\Phi d^{3}x
\end{equation}
\begin{equation}
t^{a}P_{a}=-\frac{1}{2\pi}\int(\Phi+2\vec{R}.\nabla\Phi)\nabla^{2}\Phi d^{3}x
\end{equation}
\begin{equation}
t_{A'(B}P^{A'}_{A)}=\frac{1}{\pi}\int(\vec{R}\times\nabla\Phi)\nabla^{2}\Phi d^{3}x
\end{equation}
\begin{equation}
t^{A}_{B'}\lambda^{A'B'}=\frac{i}{2\pi}\int[2\vec{R}.\nabla\Phi\vec{R}+\Phi\vec{R}-R^{2}\nabla\Phi]\nabla^{2}\Phi d^{3}x  \ .
\end{equation}
In the ML solution the conformal factor (\ref{mlform}) leads to
\begin{equation}
\nabla\Phi=\sum_{n=1}^{\infty}a_{n}\frac{\vec{d}_{n}-\vec{R}}{|\vec{R}-\vec{d}_{n}|^{3}}+b_{n}\frac{\vec{e}_{n}-\vec{R}}{|\vec{R}-\vec{e}_{n}|^{3}}
\end{equation}
and 
\begin{equation}
\nabla^{2}\Phi=-4\pi[\sum_{n=1}^{\infty}a_{n}\delta(\vec{R}-\vec{d}_{n})+
\sum_{n=1}^{\infty}b_{n}\delta(\vec{R}-\vec{e}_{n})]\ .
\end{equation}
In order to evaluate the Penrose mass of the hole 1, we choose the 2-surface
to be the sphere with center at $\vec{R}=\vec{d}_{1}$ (the throat of radius $a$).
Remembering that all the images located at $\vec{R}=\vec{d}_{2n-1}$ and $\vec{R}=\vec{e}_{2n}$ lie inside 
that sphere we obtain
\begin{eqnarray*}
& &
(\mu_{AB}t^{B}_{A'})_{1}=4i\sum_{i=1}^{\infty}\sum_{n=1}^{\infty}a_{2i-1}(a_{2n}\frac{\vec{d}_{2n}-\vec{d}_{2i-1}}{|\vec{d}_{2n}-\vec{d}_{2i-1}|^{3}}
+ b_{2n-1}\frac{\vec{e}_{2n-1}-\vec{d}_{2i-1}}{|\vec{e}_{2n-1}-\vec{d}_{2i-1}|^{3}})+\\
& &
4i\sum_{i=1}^{\infty}\sum_{n=1}^{\infty}b_{2i}(b_{2n-1}\frac{\vec{e}_{2n-1}-\vec{e}_{2i}}{|\vec{e}_{2n-1}-\vec{e}_{2i}|^{3}}+
a_{2n}\frac{\vec{d}_{2n}-\vec{e}_{2i}}{|\vec{d}_{2n}-\vec{e}_{2i}|^{3}})
\end{eqnarray*}
\begin{eqnarray*}
& &
(t^{a}P_{a})_{1}=2\sum_{i=1}^{\infty}a_{2i-1}+b_{2i}+2\sum_{i=1}^{\infty}\sum_{n=1}^{\infty}a_{2i-1}[a_{2n}\frac{d^{2}_{2n}-d^{2}_{2i-1}}{|\vec{d}_{2i-1}-\vec{d}_{2n}|^{3}}\\
& &
+b_{2n-1}\frac{e^{2}_{2n-1}-d^{2}_{2i-1}}{|\vec{d}_{2i-1}-\vec{e}_{2n-1}|^{3}}]
+2\sum_{i=1}^{\infty}\sum_{n=1}^{\infty}b_{2i}[b_{2n-1}\frac{e^{2}_{2n-1}-e^{2}_{2i}}{|\vec{e}_{2i}-\vec{e}_{2n-1}|^{3}}+a_{2n}\frac{d^{2}_{2n}-e^{2}_{2i}}{|\vec{e}_{2i}-\vec{d}_{2n}|^{3}}]
\end{eqnarray*}
\begin{eqnarray*}
[t_{A'(B}P^{A'}_{A)}]_{1}=0
\end{eqnarray*}
since all images lie along the z axis, and
\begin{eqnarray*}
& &
(t^{A}_{B'}\lambda^{A'B'})_{1}=-2i\sum_{i=1}^{\infty}a_{2i-1}\vec{d}_{2i-1}+b_{2i}\vec{e}_{2i}-2i\sum_{i=1}^{\infty}\sum_{n=1}^{\infty}a_{2i-1}[a_{2n}\frac{d^{2}_{2n}\vec{d}_{2i-1}-d^{2}_{2i-1}\vec{d}_{2n}}{|\vec{d}_{2i-1}-\vec{d}_{2n}|^{3}}\\
& &
+b_{2n-1}\frac{e^{2}_{2n-1}\vec{d}_{2i-1}-d^{2}_{2i-1}\vec{e}_{2n-1}}{|\vec{d}_{2i-1}-\vec{e}_{2n-1}|^{3}}]
-2i\sum_{i=1}^{\infty}\sum_{n=1}^{\infty}b_{2i}[b_{2n-1}\frac{e^{2}_{2n-1}\vec{e}_{2i}-e^{2}_{2i}\vec{e}_{2n-1}}{|\vec{e}_{2i}-\vec{e}_{2n-1}|^{3}}\\
& &
+a_{2n}\frac{d^{2}_{2n}\vec{e}_{2i}-e^{2}_{2i}\vec{d}_{2n}}{|\vec{e}_{2i}-\vec{d}_{2n}|^{3}}]
\end{eqnarray*}
and the Penrose quasi-local mass of black hole 1 will be
\begin{equation}
m^{2}_{1P}=[(P_{a}t^{a})_{1}]^{2}+2(\overline{\lambda}^{AB}t^{A'}_{B})_{1}(t^{C}_{A'}\mu_{AC})_{1}\ .
\end{equation}
Similarly if we take for volume of integration the sphere of radius $b$ we
get
\begin{eqnarray*}
(\mu_{AB}t^{B}_{A'})_{2}=-(\mu_{AB}t^{B}_{A'})_{1}
\end{eqnarray*}
\begin{eqnarray*}
& &
(t^{a}P_{a})_{2}=2\sum_{i=1}^{\infty}a_{2i}+b_{2i-1}-2\sum_{i=1}^{\infty}\sum_{n=1}^{\infty}a_{2i-1}[a_{2n}\frac{d^{2}_{2n}-d^{2}_{2i-1}}{|\vec{d}_{2i-1}-\vec{d}_{2n}|^{3}}\\
& &
+b_{2n-1}\frac{e^{2}_{2n-1}-d^{2}_{2i-1}}{|\vec{d}_{2i-1}-\vec{e}_{2n-1}|^{3}}]
-2\sum_{i=1}^{\infty}\sum_{n=1}^{\infty}b_{2i}[b_{2n-1}\frac{e^{2}_{2n-1}-e^{2}_{2i}}{|\vec{e}_{2i}-\vec{e}_{2n-1}|^{3}}+a_{2n}\frac{d^{2}_{2n}-e^{2}_{2i}}{|\vec{e}_{2i}-\vec{d}_{2n}|^{3}}]
\end{eqnarray*}
\begin{eqnarray*}
[t_{A'(B}P^{A'}_{A)}]_{2}=0
\end{eqnarray*}
\begin{eqnarray*}
& &
(t^{A}_{B'}\lambda^{A'B'})_{2}=-2i\sum_{i=1}^{\infty}a_{2i}\vec{d}_{2i}+b_{2i-1}\vec{e}_{2i-1}+2i\sum_{i=1}^{\infty}\sum_{n=1}^{\infty}a_{2i-1}[a_{2n}\frac{d^{2}_{2n}\vec{d}_{2i-1}-d^{2}_{2i-1}\vec{d}_{2n}}{|\vec{d}_{2i-1}-\vec{d}_{2n}|^{3}}\\
& &
+b_{2n-1}\frac{e^{2}_{2n-1}\vec{d}_{2i-1}-d^{2}_{2i-1}\vec{e}_{2n-1}}{|\vec{d}_{2i-1}-\vec{e}_{2n-1}|^{3}}]
+2i\sum_{i=1}^{\infty}\sum_{n=1}^{\infty}b_{2i}[b_{2n-1}\frac{e^{2}_{2n-1}\vec{e}_{2i}-e^{2}_{2i}\vec{e}_{2n-1}}{|\vec{e}_{2i}-\vec{e}_{2n-1}|^{3}}\\
& &
+a_{2n}\frac{d^{2}_{2n}\vec{e}_{2i}-e^{2}_{2i}\vec{d}_{2n}}{|\vec{e}_{2i}-\vec{d}_{2n}|^{3}}]
\end{eqnarray*}
and the corresponding Penrose mass will be
\begin{equation}
m^{2}_{2P}=[(P_{a}t^{a})_{2}]^{2}+2(\overline{\lambda}^{AB}t^{A'}_{B})_{2}(t^{C}_{A'}\mu_{AC})_{2}\ .
\end{equation}
To compute the total mass at the instant of time symmetry we 
choose for surface of integration a 2-sphere surrounding both throats. Then each term above add up and we get
\begin{eqnarray*}
(\mu_{AB}t^{B}_{A'})_{T}=(\mu_{AB}t^{B}_{A'})_{1}+(\mu_{AB}t^{B}_{A'})_{2}=0
\end{eqnarray*}
\begin{eqnarray*}
(t^{a}P_{a})_{T}=(t^{a}P_{a})_{1}+(t^{a}P_{a})_{2}=2\sum_{i=1}^{\infty}
a_{i}+b_{i}
\end{eqnarray*}
\begin{eqnarray*}
[t_{A'(B}P^{A'}_{A)}]_{T}=(t_{A'(B}P^{A'}_{A)})_{1}+(t_{A'(B}P^{A'}_{A)})_{2}=0
\end{eqnarray*}
and
\begin{eqnarray*}
(t^{A}_{B'}\lambda^{A'B'})_{T}=-2i\sum_{i=1}^{\infty}(a_{i}\vec{d}_{i}+b_{i}\vec{e}_{i})
\end{eqnarray*}
and hence
\begin{equation}
M=2\sum_{i=1}^{\infty}a_{i}+b_{i}
\end{equation}
which coincides with the ADM mass (appendix B, expression (\ref{totalmass})). As expected, this mass is not equal to the sum of the two individual masses. 
The difference is the binding energy
\begin{equation}
E_{B}=M-m_{1P}-m_{2P}\ .
\end{equation}
In the special case in which the holes have equal mass, we obtain
\begin{equation}
(t^{a}P_{a})_{1}=(t^{a}P_{a})_{2}=\textstyle\frac{1}{2}M
\end{equation}
\begin{equation}
(\mu_{AB}t^{B}_{A'})_{1}=-(\mu_{AB}t^{B}_{A'})_{2}=-iI=-4i\sum_{n=1}^{\infty}
\sum_{m=1}^{\infty}\frac{\sinh n\mu_{0}\sinh m\mu_{0}}{\sinh^{2}(n+m)\mu_{0}}
\end{equation}
\begin{eqnarray*}
& &
(t^{A}_{B'}\lambda^{A'B'})_{1}=-(t^{A}_{B'}\lambda^{A'B'})_{2}=-iJ=
-2ia^{2}\sinh \mu_{0}(\sum_{n=1}^{\infty}\frac{\cosh n\mu_{0}}{\sinh^{2}n\mu_{0}}\\
& &
+\sum_{n=1}^{\infty}\sum_{m=1}^{\infty}\frac{\cosh n\mu_{0}\cosh m\mu_{0}}{\sinh^{2}(m+n)\mu_{0}})
\end{eqnarray*}
giving 
\begin{equation}
m^{2}_{1P}=m^{2}_{2P}=\textstyle\frac{1}{4}M^{2}+2IJ
\end{equation}
a result obtained by Tod\cite{tod}.


\begin{thebibliography}{10}

\bibitem{ligo} A.~A.~Abramovici {\em et al.}, Science {\bf 256}, 325
(1992); K.~S.~Thorne, submitted to {\it Proceedings of Snowmass 94
Summer Study on Particle and Nuclear Astrophysics and Cosmology},
eds. W.~W.~Kolb and R.~Peccei (World Scientific, Singapore).

\bibitem{grandchallenge} Proceedings of the November 1994 meeting
of the Grand Challenge Alliance to study black hole collisions may be
obtained by contacting E.~Seidel at NCSA (unpublished).

\bibitem{AP1}A.~Abrahams and R.~H.~Price, Phys.\ Rev.\ {\bf D53}, 1963 (1996).

\bibitem{price_pullin94}
R.~H. Price and J.~Pullin, Phys.\ Rev.\ Lett.\ {\bf 72}, 3297 (1994).

\bibitem{all}P.~Anninos, R.~H.~Price, J.~Pullin, E.~Seidel and W.-M.~Suen, Phys.\ Rev.\ D {\bf 52}, 4462 (1995).

\bibitem{AP2}A.~Abrahams and R.~H.~Price, Phys.\ Rev.\ {\bf D53}, 1972 (1996).

\bibitem{abrahams_cook94}
A.~M. Abrahams and G.~B. Cook, Phys.\ Rev.\ D
{\bf50}, R2364 (1994).

\bibitem{baker_boost}J.~Baker, A.~Abrahams, P.~Anninos, S.~Brandt, R.~H.~Price,
J.~Pullin, E.~Seidel, preprint gr-qc/9608064 (1996) .

\bibitem{screw} H.-P.~Nollert, private communication.

\bibitem{2ndorder}R.~Gleiser, C.~O.~Nicasio, R.~H.~Price, and
  J.~Pullin, Class. Quantum Grav. {\bf 13} L117 (1996).

\bibitem{prep}R.~Gleiser, C.~O.~Nicasio, R.~H.~Price, and J.~Pullin,
  to appear in Phys.\ Rev.\ Lett.

\bibitem{BowenYork}J.~Bowen and J.~W.~York, Jr., Phys. Rev. 
{\bf D21}, 2047 (1980).

\bibitem{bl}D. R. Brill and R. W. Lindquist, Phys.\ Rev. {\bf131}, 471 (1964).

\bibitem{images}C. Misner, Ann.~Phys. (N.Y.) {\bf 24}, 102 (1963).

\bibitem{misner}C. Misner, Phys. Rev. {\bf 118}, 1110 (1960).

\bibitem{lind63} R.\ W.\ Lindquist, J. Math. Phys., {\bf 4}, 938 (1963).

\bibitem{penrose}R. Penrose, Proc. R. Soc. Lond. {\bf A 381}, 53 (1982).

\bibitem{tod}K. P. Tod,  Proc. R. Soc. Lond. {\bf A 388}, 457 (1983).

\bibitem{global}R.~Penrose, in {\em Global Riemannian Geometry},
edited by T. J. Willmore and N. Hitchin (Ellis Horwood Limited, Chichester,
England, 1984), p. 203.

\bibitem{moncrief74}
V. Moncrief, Ann. Phys. (NY) {\bf 88},  323  (1974).
\bibitem{zerilli} F.~Zerilli, Phys.\ Rev.\ Lett. {\bf 24} 737, (1971).

\bibitem{loustoprice}C.\ O.\ Lousto and R.\ H.\ Price, preprint
  gr-qc/9609012(1996).

\bibitem{moncrief80}V. Moncrief, Ap. J. {\bf 238},
333 (1980).
\bibitem{Suen} W.-M.~Suen, private communication.

\bibitem{smythe} William R. Smythe, {\em Static and dynamic electricity} (McGraw Hill, New York, 1968).

\bibitem{bishop} N.~T.~Bishop, Gen. Rel. Grav. {\bf 16}, 589 (1984).

\end{thebibliography}
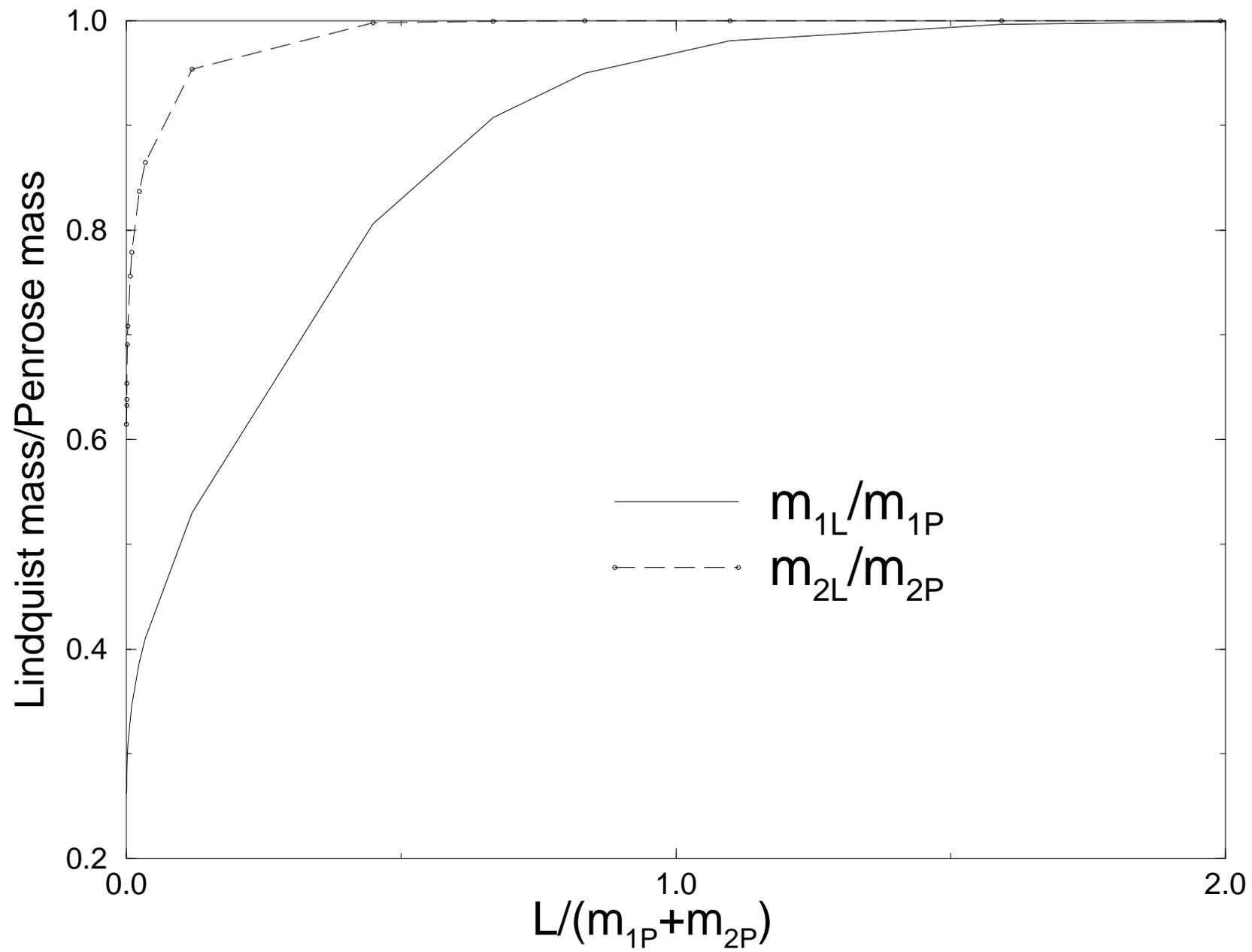
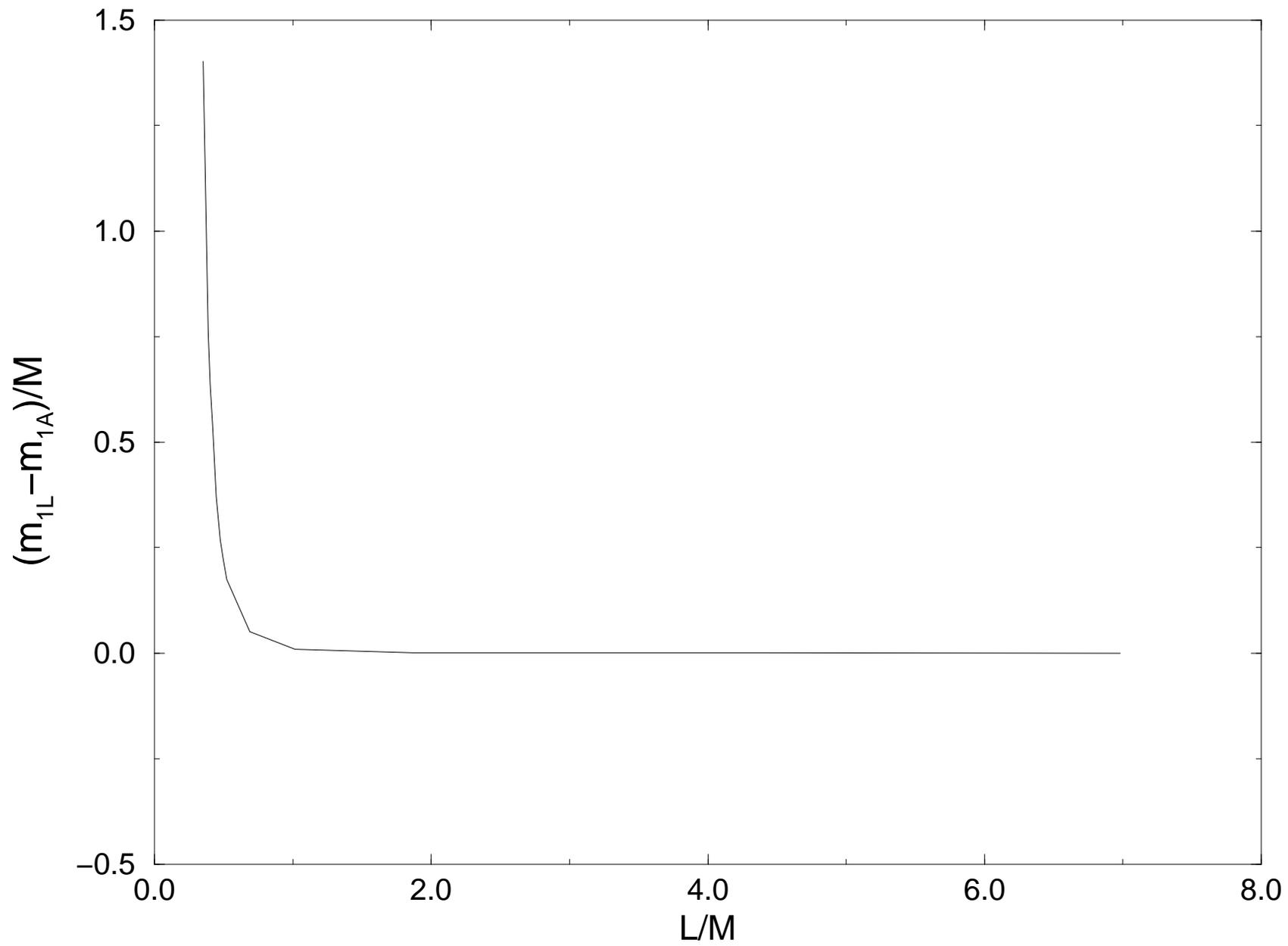
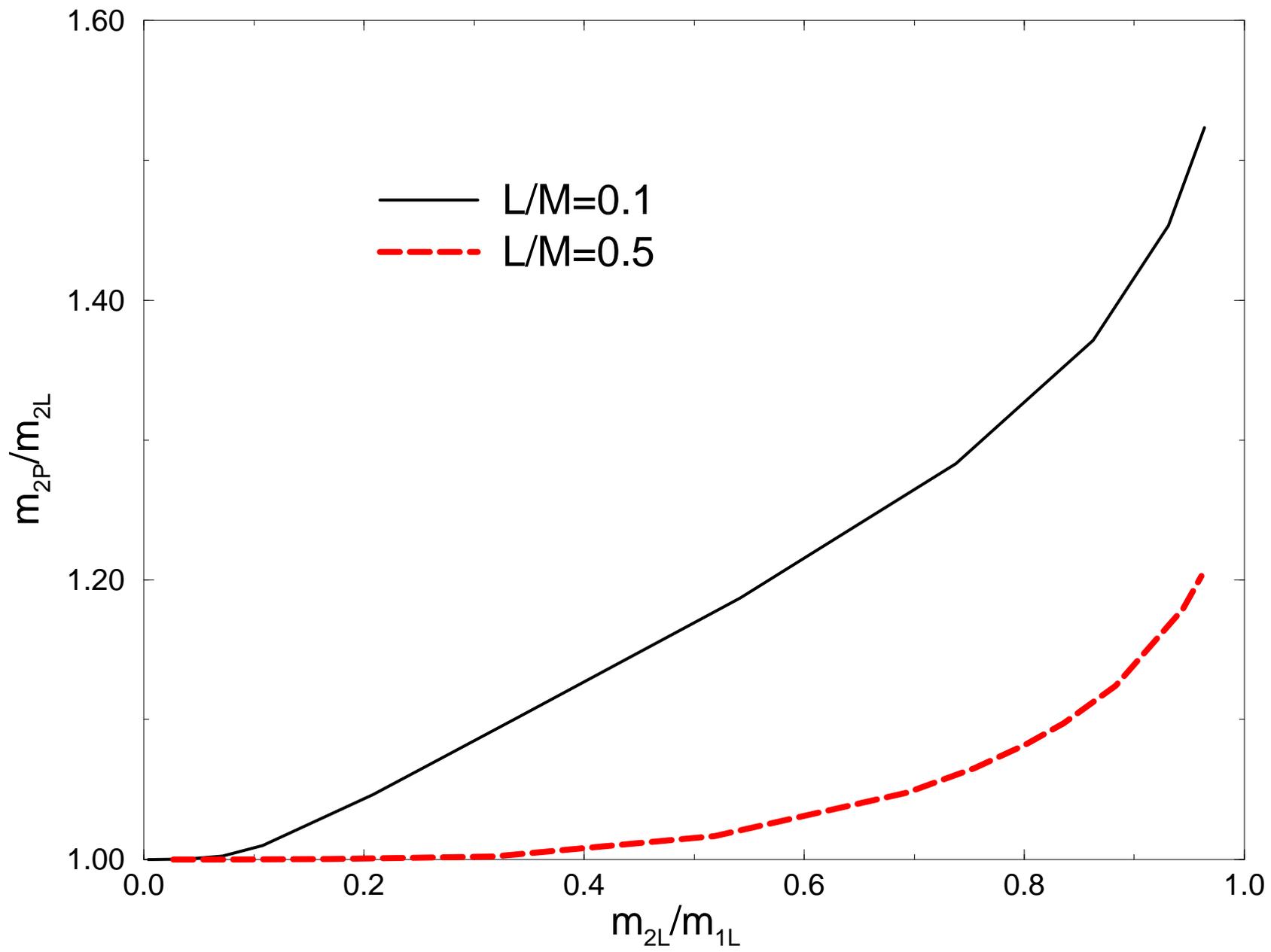
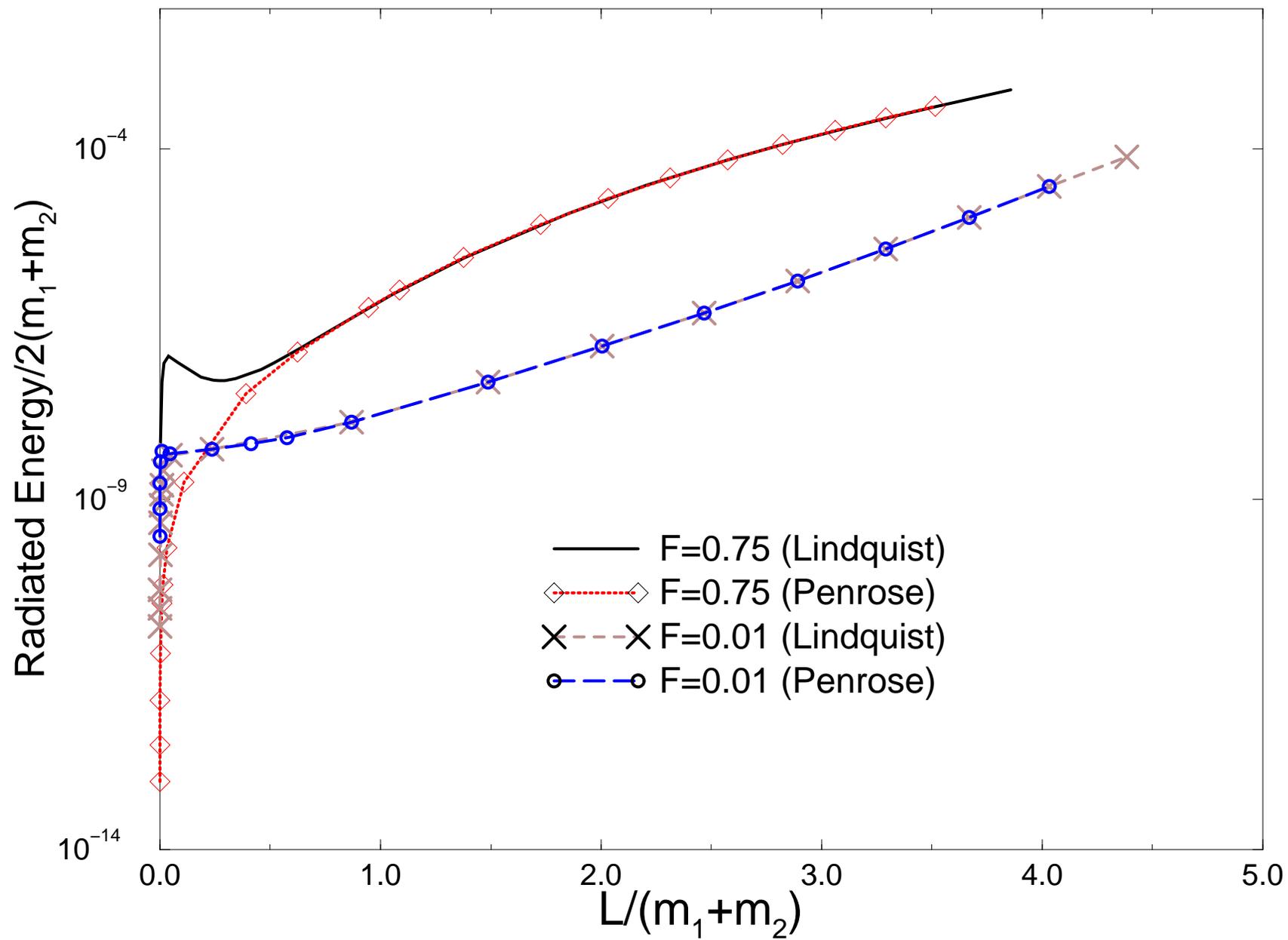
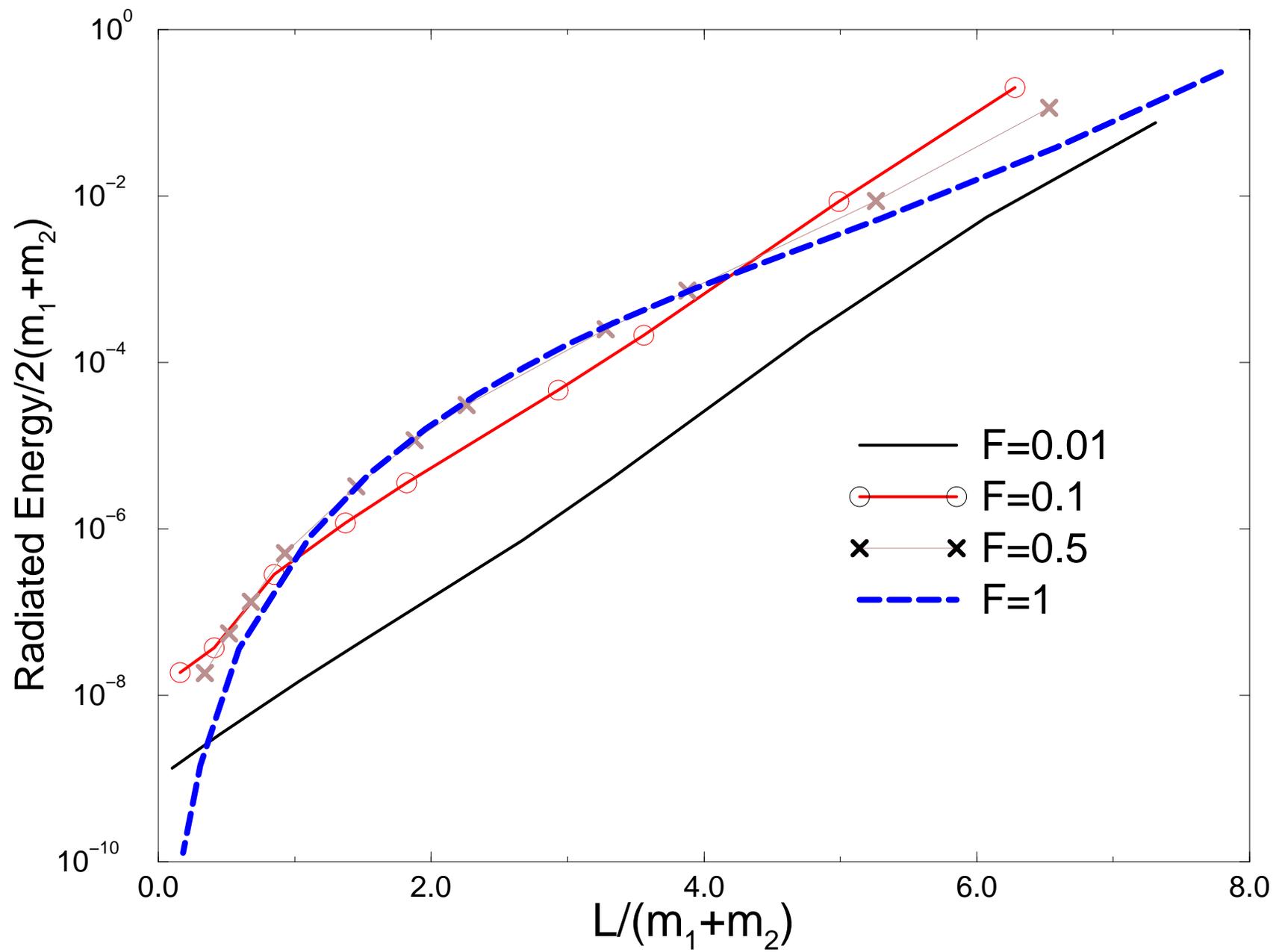
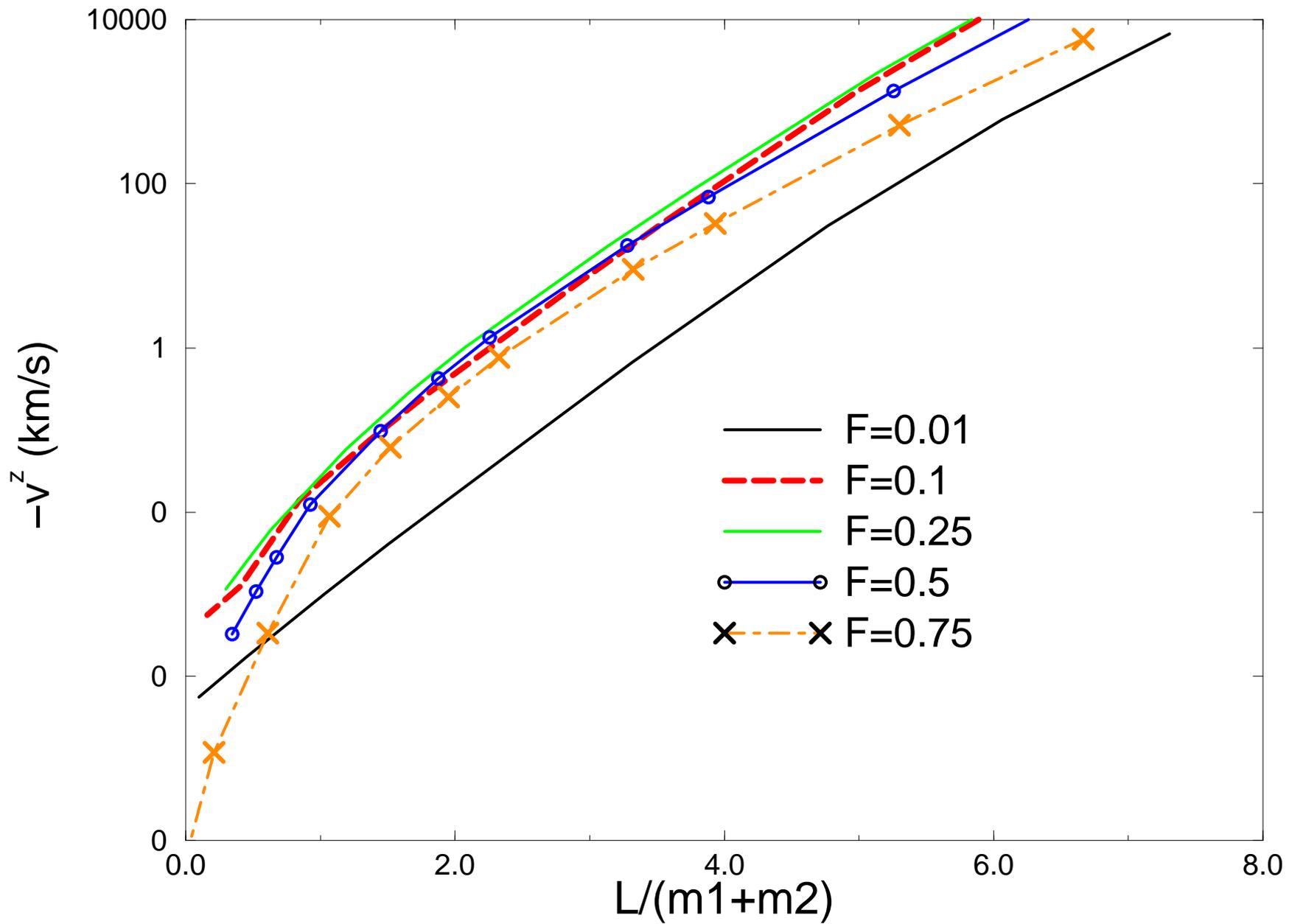
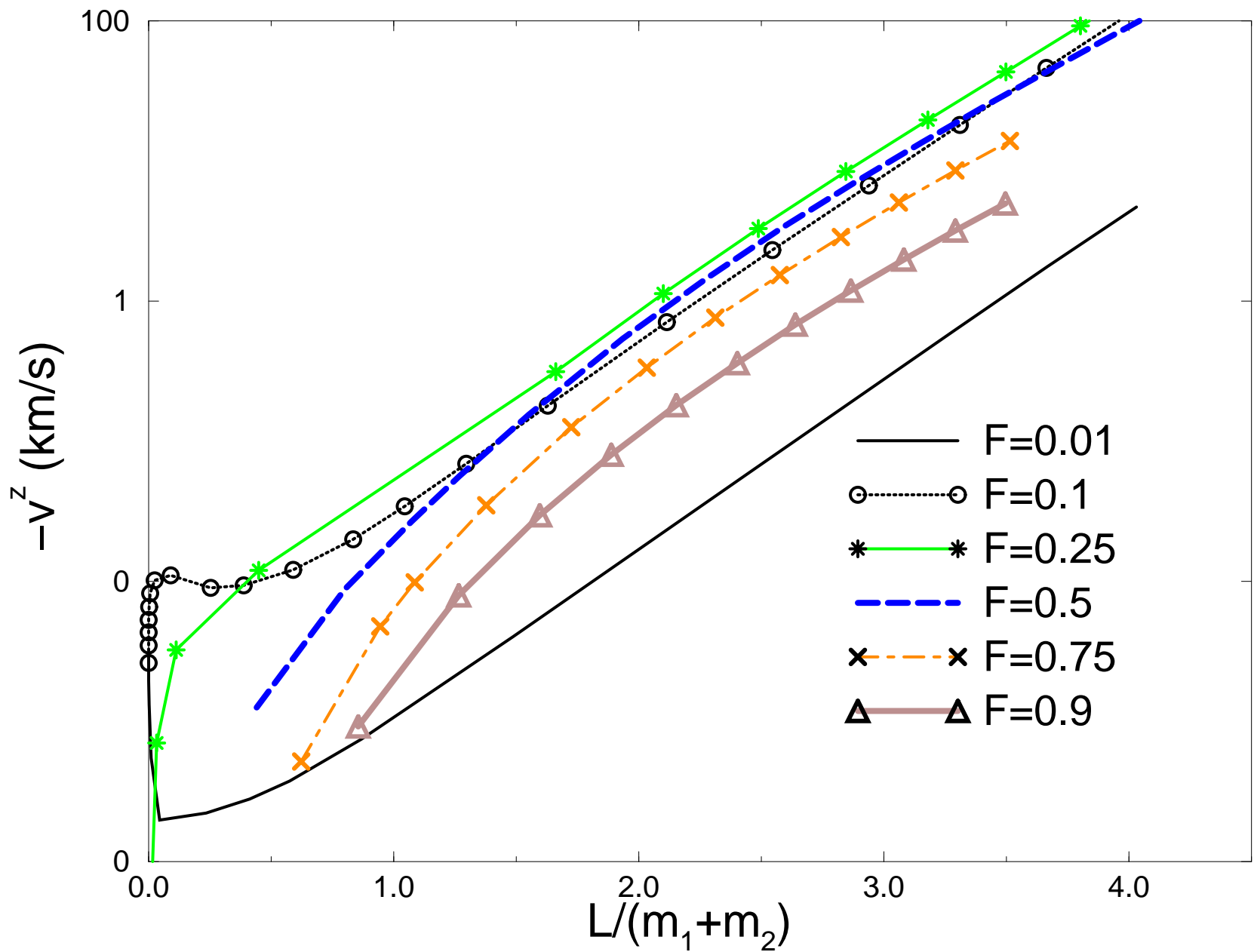
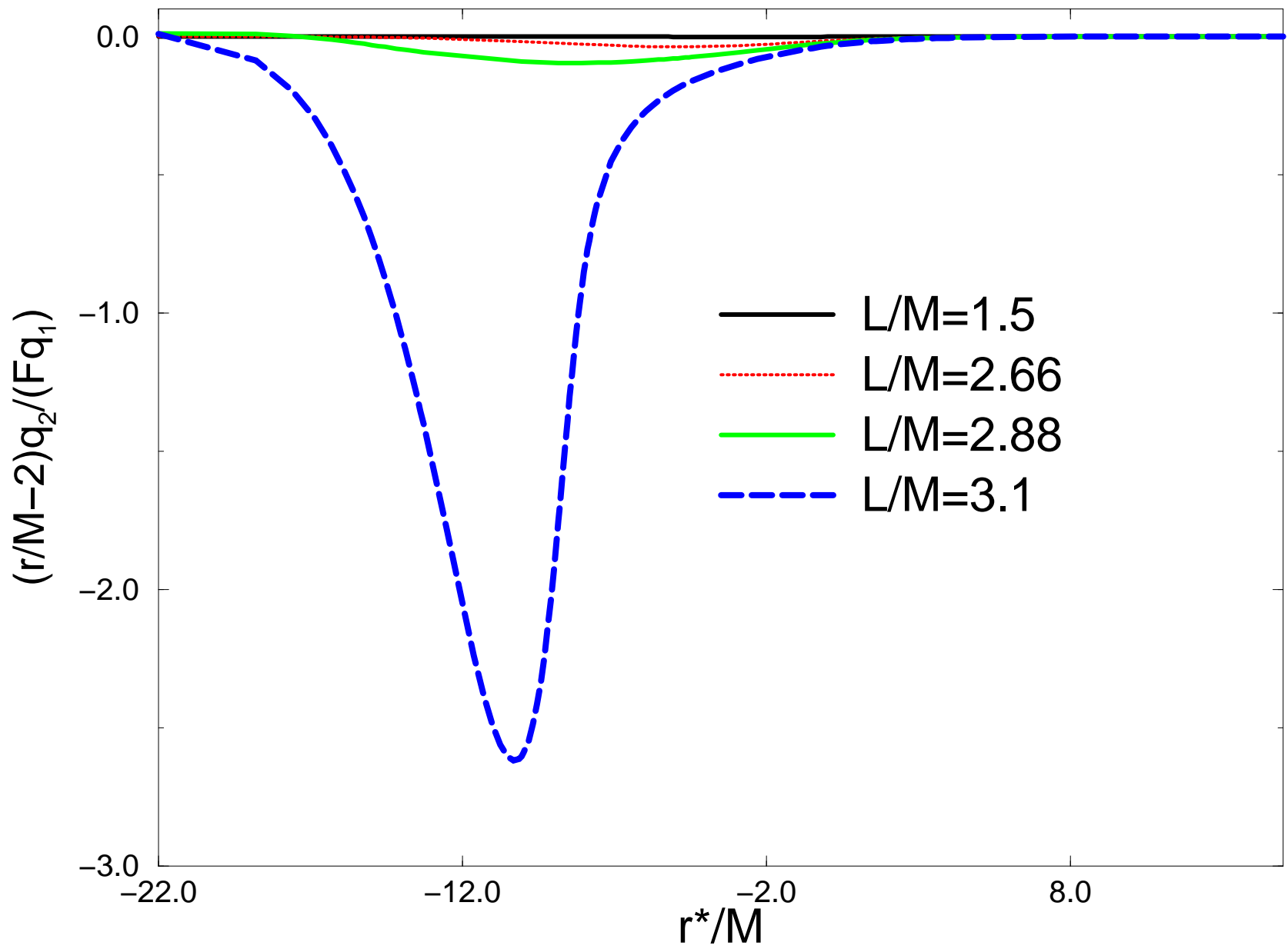
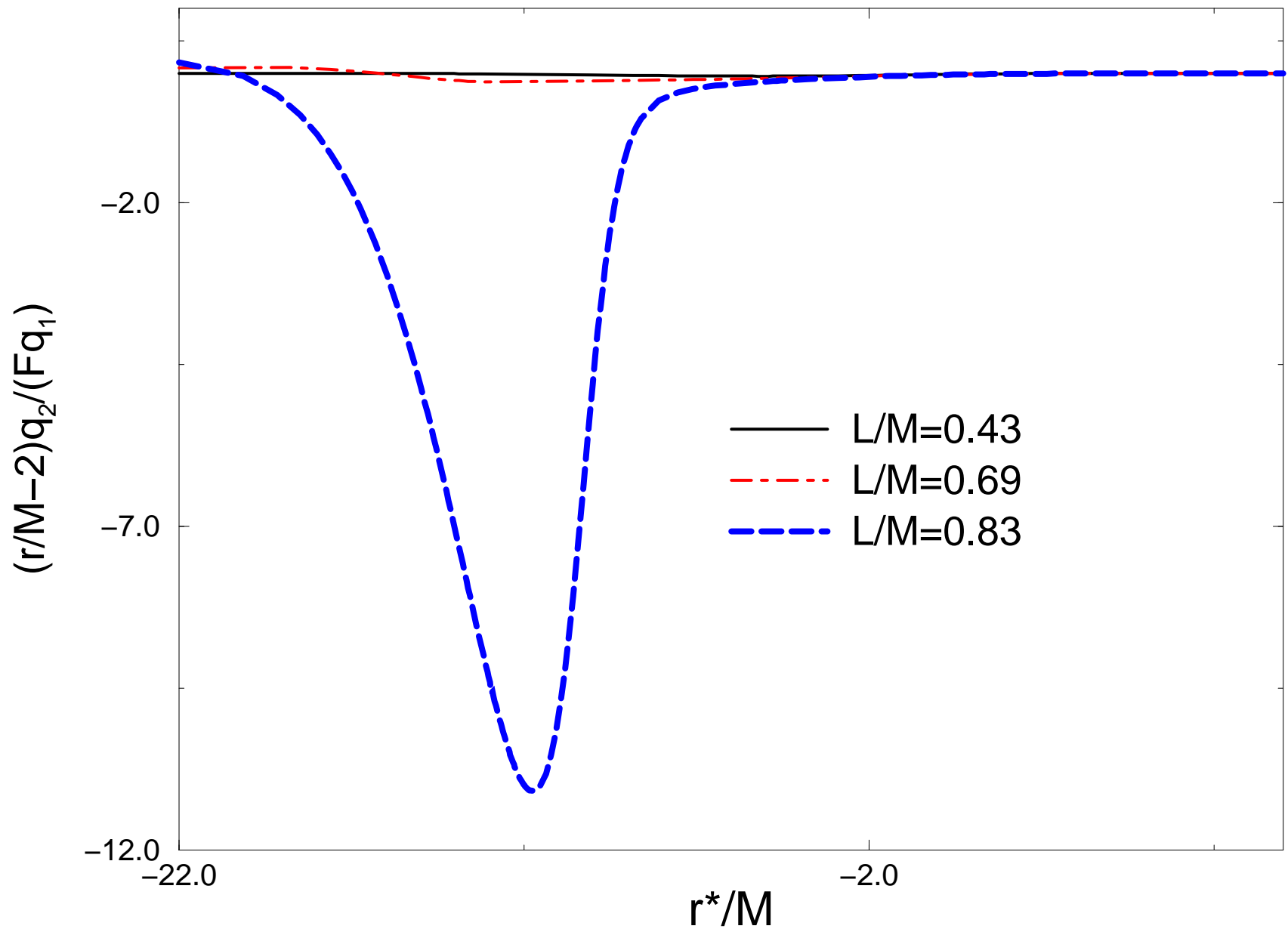
\end{document}